\documentclass[a4paper,fleqn,usenatbib]{mnras}

\usepackage{newtxtext,newtxmath}
\usepackage[T1]{fontenc}
\usepackage{ae,aecompl}
\usepackage{color}

\usepackage{graphicx}
\usepackage{amsmath}
\usepackage{amssymb}
\usepackage{multirow}
\usepackage{xspace}

\newcommand{\ie}{i.e.\xspace}
\newcommand{\eg}{e.g.\xspace}

\newcommand{\Msun}{{\rm M}_{\odot}}
\newcommand{\Lsun}{{\rm L}_{\odot}}
\newcommand{\Rsun}{{\rm R}_{\odot}}

\newcommand{\Teff}{T_{\rm eff}}
\newcommand{\Rc}{R_{\rm c}}
\newcommand{\Mc}{M_{\rm c}}
\newcommand{\aml}{\alpha_{\rm ML}}
\newcommand{\numax}{\nu_{\rm max}}
\newcommand{\nuac}{\nu_{\rm ac}}

\newcommand{\ks}{K_{\rm s}}

\newcommand{\Zref}{Z_{\rm ref}}
\newcommand{\xc}{X_{\rm C}}
\newcommand{\xn}{X_{\rm N}}
\newcommand{\xo}{X_{\rm O}}
\newcommand{\co}{{\rm C}/{\rm O}}
\newcommand{\fco}{f_{\co}}

\title[New O-rich and C-rich models of LPVs]
{Modelling Long-Period Variables -- I.\\A new grid of O-rich and C-rich pulsation models}

\author[M. Trabucchi et al.]{
Michele Trabucchi,$^{1}$\thanks{Contact e-mail: \href{mailto:michele.trabucchi@unipd.it}{michele.trabucchi@unipd.it}}
Peter R. Wood,$^{2}$
Josefina Montalb{\'a}n,$^{1}$
\newauthor
Paola Marigo,$^{1}$
Giada Pastorelli$^{1}$
and
and L{\'e}o Girardi$^{3}$
\\
$^{1}$Dipartimento di Fisica e Astronomia, Universit\`a di Padova,
Vicolo dell'Osservatorio 2, I-35122 Padova, Italy \\
$^{2}$Research School of Astronomy and Astrophysics,
Australian National University, Canberra, ACT2611, Australia \\
$^{3}$Osservatorio Astronomico di Padova - INAF, Vicolo dell'Osservatorio 5, I-35122 Padova, Italy
}

\date{Accepted XXX. Received YYY; in original form ZZZ}

\pubyear{2018}

\begin{document}
\label{firstpage}
\pagerange{\pageref{firstpage}--\pageref{lastpage}}
\maketitle

\begin{abstract}
    We present a new grid of non-adiabatic, linear pulsation models
    of Long-Period Variables (LPVs), including periods and growth rates for
    radial modes from the fundamental to the fourth overtone. The
    models span a wide range in mass, luminosity, metallicity, $\co$
    ratio and helium abundance, effectively covering the whole thermally-pulsing asymptotic
    giant branch (TP-AGB) evolution, and representing a significant update with
    respect to previous works. The main improvement is the inclusion
    of detailed atomic and molecular opacities, consistent with the models
    chemical mixture, that makes the present set of models the
    first to systematically account for variability in C-stars. We examine
    periods and growth rates in the models, and find that,
    while the fundamental mode is affected by the structure of
    the envelope, overtones are less sensitive to the interior and
    largely determined by the global properties. In the models, the
    frequency of the overtone with the largest degree of excitation
    is found to scale with the acoustic cut-off frequency at
    the stellar surface, a behaviour similar to that observed for
    the frequency of maximum oscillation power for solar-like oscillations in
    less evolved red giants. This allows us to provide a
    simple analytic prescription to predict the most-likely dominant mode as
    a function of stellar parameters. Best-fit relations for periods are
    also provided. By applying results of pulsation models to evolutionary
    tracks, we present a general picture of the evolution of
    long-period variability during the TP-AGB, that we find consistent with
    observations. Models are made public through a dedicated web interface.
\end{abstract}


%%%%%%%%%%%%%%%%%%%%%%%%%%%%%%%%%%%%%%%%%%%%%%%%%%

%%%%%%%%%%%%%%%%% BODY OF PAPER %%%%%%%%%%%%%%%%%%

\section{Introduction}
\label{sec:Introduction}

The thermally-pulsing asymptotic giant branch (TP-AGB) is the final
stage in the evolution of low- and intermediate-mass stars.
Among the plethora of complex processes involved in this
evolutionary phase, stellar pulsation is one of the most interesting.
Not only it is intimately interconnected with poorly
understood processes such as dust formation and mass loss
\citep[see, \eg, the recent review by][]{Hofner_Olofsson_2018},
but it is also an extremely valuable observable for
the calibration of theoretical models and the estimate
of global stellar parameters, since pulsation periods are strongly linked to
stellar mass, radius and effective temperature.
Owing to their large radii, these stars exhibit variability
characterised by periods of order of hundreds of days,
hence the name of long-period variables (LPVs).
The brightest and largest-amplitude ones, the Mira variables,
are observable at large distances, especially at infrared wavelengths.
In that spectral range, they follow a clear period-luminosity (PL) relation,
which makes them very promising distance indicators
\citep{Whitelock_etal_2013,Whitelock_Feast_2014,Menzies_etal_2015,Yuan_etal_2017,Huang_etal_2018}.
The discovery of several other PL sequences in the LPV population
of the Large Magellanic Cloud (\citealt{Wood_etal_1999}; see also
\citealt{Wood_2000,Soszynski_etal_2009_LMC,Wood_2015,Trabucchi_etal_2017}), re-ignited
interest in such stars. The additional sequences are populated by
Semi-regular variables (SRVs) as well as OGLE Small Amplitude Red Giants
\citep[OSARGs;][]{Wray_etal_2004,Soszynski_etal_2004_OSARG},
a relatively recent variability type associated with both
red giant branch (RGB) and AGB stars, that make up for the majority of LPVs.
SRVs and OSARGs oscillate (often simultaneously)
in multiple low-order radial and non-radial modes \citep{Soszynski_etal_2004_OSARG},
characterised by distinct periods and excitation properties
which depend on the stellar parameters of the evolutionary stage.
They also have interesting properties which can be used
for distance measurements \citep{Rau_etal_2018}.
Overall, LPVs span several orders of magnitudes in each of
luminosity (from $\sim10^2$ to $\sim10^5\,\Lsun$),
period (from a few days to a few thousands days) and amplitude
(from $\sim10^{-3}$ mag to several magnitudes), reflecting
the dramatic changes in the structure of red giants during
their late evolution. Their observed periods and amplitudes
provide additional constraints, together with other observables,
to be matched by models, allowing us to refine
our knowledge of stellar structure and evolution.

Currently, the interpretation of LPVs
in the framework of evolutionary models often relies on analytic prescriptions derived from
restricted grids of pulsation models
\citep[\eg,][]{Fox_Wood_1982,Wood_etal_1983,Ostlie_Cox_1986,Wood_1990},
whereas updated models of luminous red giant variables
have long been missing from the scientific literature,
and a set of models systematically accounting for the
full variety of TP-AGB stars has never been published.

Some recent advancement in this field include
the work by \citet{Xiong_etal_2018} \citep[see also][]{Xiong_Deng_2007,Xiong_Deng_2013},
who addressed the study of pulsation stability
in red giants using a theory of non-local,
time-dependent convection, and the promising
results of 3D radiation-hydrodynamic models
of \citet[][and references therein]{Freytag_etal_2017}.
However, the time-consuming nature
of such models makes them unsuitable
for the construction of large grids.

A critical shortcoming of previous models, with the
exception of a few selected studies, is that they
do not account for surface chemical enrichment expected in TP-AGB phase.
C-type stars, produced by the dredge-up of carbon,
have characteristic spectral features that are
dramatically different from those of their O-rich counterparts.
This is a consequence of altered molecular equilibria,
and the corresponding drastic change in the main
sources of molecular opacity. Since atmospheric
opacities affect stellar radii, it is reasonable
to expect pulsation periods to be affected as well.
Yet, whether C-rich and O-rich LPVs follow different
PL relations has never been thoroughly addressed theoretically.
Also, the dependence of PL relations upon metallicity
is poorly studied from a theoretical point of view,
despite the metallicity dependence being crucial for
the calibration of PL relations as standard candles.

In this work, we present a new, large grid of linear, radial,
non-adiabatic pulsation models of luminous red giants,
with updated opacity data for CNO-varied metal mixtures.
It was designed to widely cover the space of stellar
parameters characterising the AGB evolution in terms of
total mass, core mass, luminosity, effective temperature
and chemical composition. Models include periods and
amplitude growth rates for five radial pulsation modes,
from the fundamental to the fourth overtone.
Growth rates allow us to predict stability/instability of individual modes,
and are used as a proxy of the expected observability.
We use results from these models to discuss the evolution
and properties of pulsation in LPVs.

This paper is structured as follows.
In Sect.~\ref{sec:ModelsMethods} we describe
the relevant aspects in the modelling of static envelopes
and the grid coverage of stellar parameters.
We then discuss the stability analysis of the envelopes and their pulsation properties.
In Sect.~\ref{sec:Stability} and~\ref{sec:Periods},
we make use of selected models in the grid to present
a simplified picture of the evolution of long-period
variables in terms of pulsational stability and of
oscillation periods. We expand the analysis to
evolutionary tracks in Sect.~\ref{sec:PulsationEvoTrack},
while Sect.~\ref{sec:Conclusions} is dedicated to
a summary and to conclusions.

\section{Models and methods}
\label{sec:ModelsMethods}

The grid of pulsation models presented here
was constructed with codes described
in \citet{Wood_Olivier_2014}, based on those
presented in \citet{Fox_Wood_1982}.
The computation involves two steps:
(1) the construction of a spherically
symmetric, static envelope  (Sect.~\ref{sec:StaticModels}), and
(2) the linear stability analysis of
its structure (Sect.~\ref{sec:PulsationModels}).

\subsection{Static envelope models}
\label{sec:StaticModels}

We computed spherically symmetric, static envelope models
covering the range of stellar parameters necessary to study and interpret the
observed pulsation properties of TP-AGB stars.

The computation of a static model
is limited to the outer envelope
(the layers above the H-burning shell).
This simplifying assumption is justified
by the fact that the stellar core and
the envelope are effectively decoupled
from a dynamical point of view, so
that radial pulsation is essentially an envelope process.
The value of the core mass, provided as
an inner boundary condition, is computed from
a core mass-luminosity relation derived
from evolutionary models, while the core
radius is kept fixed at $\Rc=0.15\,\Rsun$
as its effect on pulsation is negligible
for the models considered here (see Sect.~\ref{sec:CoreMassRadius}).
Additional boundary conditions include
the values of total (current) mass,
luminosity and chemical composition.
The latter is defined by the metallicity ($Z$)
and by the mass fractions of hydrogen ($X$),
helium ($Y$), carbon ($\xc$), nitrogen ($\xn$)
and oxygen ($\xo$).
Convective energy transport is described
by means of the mixing length theory \citep{Cox_Giuli_PSS_1968}
in both static and perturbed models.

\subsection{Opacity data}
\label{sec:OpacityData}

Third dredge-up events are able to produce
significant changes in the surface composition
of TP-AGB stars \citep[see, \eg,][]{Lattanzio_Wood_2003}.
When sufficient amounts of carbon
produced by nuclear reactions in the interior
are mixed into the envelope, the abundance of
carbon by number can exceed that of oxygen,
initially larger, leading to the formation of carbon stars.
This is usually described in terms of
the number ratio $\co$ of carbon and oxygen atoms at the surface.

The low effective temperature of TP-AGB stars
favours the formation of molecules,
and the CO molecule, due to
its high binding energy, effectively locks all
atoms of the least abundant of the elements C and O.
All the remaining atoms of the most abundant of
those two elements can then form other molecules.
Therefore, molecular equilibria exhibit an abrupt
change in the vicinity of the transition at $\co\sim1$,
which is mirrored by a drastic alteration of
the main sources of molecular opacity.
This is very evident in the spectral energy
distribution of O-rich stars, dominated by absorption
bands of TiO, VO, H$_2$O, while
C-rich star exhibit strong absorption features
of carbon-bearing molecules such as C$_2$, CN and SiC.

At a given stellar luminosity, larger atmospheric opacities
correspond to lower effective temperatures
and larger radii, \ie, lower mean densities.
Since pulsation periods depend on the mean density $\bar\rho$
of the envelope, it is clear that chemically-consistent
opacity data are necessary for the appropriate
modelling of C-rich variable stars.
For this reason, we employed detailed atomic and
molecular opacities computed with the \texttt{{\AE}SOPUS}
code \citep{Marigo_Aringer_2009} in the pulsation models.
Such Rosseland mean opacities are appropriate
to describe the low-temperature range corresponding
to the outer layers of the envelope ($3.2\leq\log(T[{\rm K}])\leq4.5$).
For the high-temperature regime (\ie, the stellar interior)
we used the data from the Opacity Project
\citep[\texttt{OP},][]{Seaton_2005},
covering the range $3.5\leq\log(T[{\rm K}])\leq8.0$.
For each combination of metallicity, hydrogen
abundance, and $\co$ ratio  (see Table~\ref{tab:PulsationGridNodes}
and Sect.~\ref{sec:ChemicalComposition}), we computed
two tables of opacity as a function
of temperature and density, one with
\texttt{{\AE}SOPUS} and the other with the \texttt{OP} routines.
Each pair of tables was then smoothly
merged in the overlapping temperature range.

\subsection{Coverage of stellar parameters}
\label{sec:CoverageStellarParameters}

\begin{table}
\normalsize
\centering
\caption{
    Parameters and nodes defining the structure of the grid.
    }
\label{tab:PulsationGridNodes}
\begin{tabular}{c|ccccc}
Parameter & \multicolumn{5}{c}{Nodes} \\
\hline
\hline
\multirow{2}{*}{$M/\Msun$} & \multicolumn{5}{c}{[0.6, 1.0] (step: 0.05), [1.0, 2.0] (step: 0.1)} \\
&  \multicolumn{5}{c}{and [2.0, 7.0] (step: 0.2)} \\
\hline
\multirow{2}{*}{$\log(L/\Lsun)$} & \multicolumn{5}{c}{[2.5, 5.0], step: 0.01} \\
& \multicolumn{5}{c}{\small (boundaries depend on $M$, see Sect.~\ref{sec:MassLuminosity})} \\
\hline
\multirow{2}{*}{$\Mc/\Msun$} & \multicolumn{5}{c}{Defined as a function} \\
& \multicolumn{5}{c}{of luminosity (see Eq.~\ref{eq:CoreMassLuminosityFunctions} and Table~\ref{tab:McupMclowCoefficients})}\\
\hline
$\aml$ & & 1.5, & 2.0, & 2.5 & \\
\hline
\multirow{2}{*}{$\Zref$} & 0.001, & 0.002, & 0.004, & 0.006, & 0.008, \\
& & 0.01, & 0.014, & 0.017 & \\
\hline
$X$    & & 0.6, & 0.7, & 0.8 & \\
\hline
\multirow{2}{*}{$\co$} & 0.3, & 0.55, & 0.7, & 0.95, & 1.0, \\
& 1.05, & 1.3, & 1.8, & 3.0, & 5.0 \\
\hline
\multirow{2}{*}{$\fco$} & $-$0.263, & 0.0, & 0.105, & 0.238, & 0.260, \\
& 0.281, & 0.374, & 0.515, & 0.737, & 0.959 \\
\hline
\end{tabular}
\end{table}

We constructed the grid of pulsation models by
varying several input parameters for the computation of
static envelope models.
To select the relevant parameters to be varied,
as well as the corresponding range and sampling,
we made use of detailed TP-AGB evolutionary tracks
computed with the \texttt{COLIBRI} code \citep{Marigo_etal_2013, Marigo_etal_2017}.
The grid coverage is summarised in Table~\ref{tab:PulsationGridNodes}.

\subsubsection{Chemical composition}
\label{sec:ChemicalComposition}

In the computation of the present models,
we take as reference the solar composition by \citet{Caffau_etal_2011}.
First, we computed a set of envelope models
of scaled-solar composition, with metallicity
$Z$ from $0.001$ to $0.017$, and hydrogen mass fraction $X$:  $0.6$, $0.7$, and $0.8$.
For each combination of $X$ and $Z$ we then computed
models with altered carbon abundance, described
by the carbon variation factor\footnote{
    Logarithms are to base 10 throughout this paper.
}
\citep[see also][]{Marigo_Aringer_2009} defined as
\begin{equation}
    \fco = \log\left(\frac{\mathrm{C}}{\mathrm{O}}\right) - \log\left(\frac{\mathrm{C}}{\mathrm{O}}\right)_{\rm ref.} \,,
\end{equation}
where C and O are the abundances by number of carbon and oxygen,
respectively, and $(\co)_{\rm ref.}=(\co)_{\odot}=0.54954$,
or alternatively
\begin{equation}
    \fco = \log\left(\frac{Y_{\rm C}}{Y_{\rm O}}\right) - \log\left(\frac{Y_{\rm C}}{Y_{\rm O}}\right)_{\rm ref.} \,,
\end{equation}
where $Y_{\rm C}=\xc/A_{\rm C}$, $Y_{\rm O}=\xo/A_{\rm O}$,
$\xc$ and $\xo$ are the abundances of carbon and oxygen
by mass fraction, and $A_{\rm C}$, $A_{\rm O}$ are the corresponding
mass numbers.
The grid ranges from $\fco=-0.263$ to $\fco=0.959$,
corresponding to $\co$ ratios from $0.3$ (sub-solar) to $5.0$
(see Table~\ref{tab:PulsationGridNodes}).
Note that, with the exception of the scaled-solar case ($\fco=0$, $\co=(\co)_{\odot}$),
metallicity is not preserved when altering the abundance of carbon.
For this reason, it is useful to identify models
in terms of their `reference metallicity' $\Zref$,
\ie, the metallicity they would have if no carbon
variation was applied. The reference metallicity
can be thought as that at the beginning of the AGB,
\ie, prior to any third dredge-up event
(see also Appendix~\ref{sec:InterpolationRoutine}).

We performed test calculations, over a limited portion
of the space of parameters, in which variations
of $\co$ have been obtained by changing the individual
abundances of carbon and/or oxygen. The results suggest
that the abundances of C and O, individually, have a minor
effect on pulsation properties, while it is really
$\co$ which plays a significant role.

Therefore, the assumption was made that variations of $\co$ are entirely
due to changes in the amount of carbon, with oxygen abundance
being always kept fixed to the scaled-solar value.
The effect of changing the envelope abundance of oxygen,
as well as an expansion of the metallicity range,
will be addressed in a future version of the present grid of models.

\begin{figure}
    \includegraphics[width=\columnwidth]{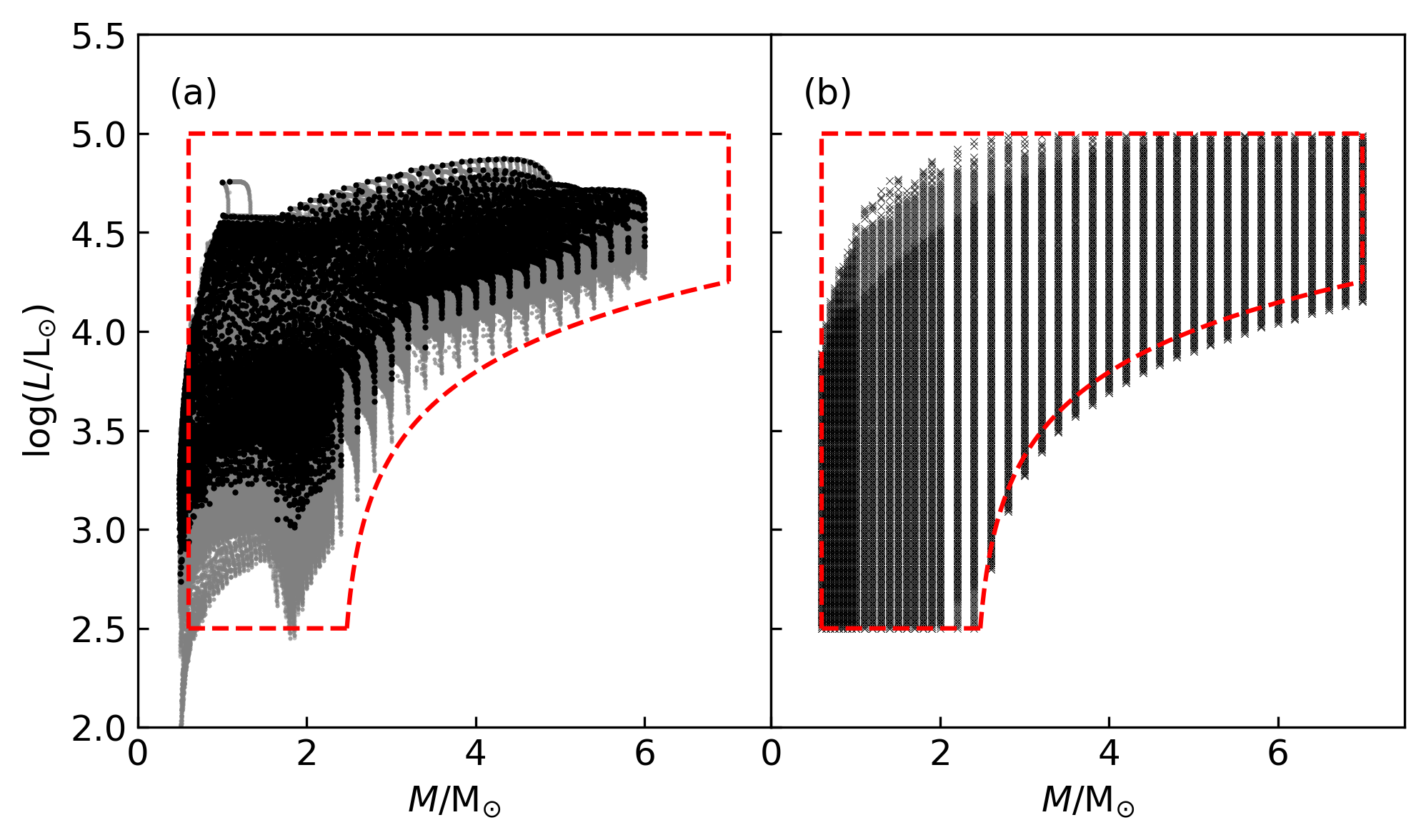}
    \caption{TP-AGB evolutionary models from \texttt{COLIBRI} (left panel)
    and pulsation models from this work (right panel)
    in the mass-luminosity plane. Evolutionary tracks
    including the whole thermal pulse cycle are plotted
    in grey colour, while black symbols represent quiescent
    points only (when the He-burning shell is dormant
    and energy is produced by the H-burning shell in quiet conditions).
    Dashed contours in both panels represent the nominal coverage of the grid.
    In some cases, the starting luminosity for a series of models
    was set slightly below that defined by Eq.~\ref{eq:LowLumFunction}
    in order to facilitate calculations. As a result, models go below
    the nominal coverage in the higher-mass range of the diagram.}
    \label{fig:MassLuminosityCoverage}
\end{figure}

\subsubsection{Mass and luminosity}
\label{sec:MassLuminosity}

We computed pulsation models with masses in
the range $0.6\leq M/\Msun \leq 7.0$. Note that
such values represent the current mass, and not the initial one.
TP-AGB stars cover several orders of magnitudes of
bolometric luminosity. Our models can have a luminosity
in the interval $2.5\leq\log(L/\Lsun)\leq5.0$ (Fig.~\ref{fig:MassLuminosityCoverage}),
the actual range depending on mass, and to a minor extent
on other input parameters. This interval covers the
whole AGB, and includes the brightest part of the RGB.

To account for the fact that the more massive
stars begin the TP-AGB phase at a higher luminosity,
we set the lower luminosity limit $L_\mathrm{low}$ as a function of
mass, according to the following expression derived
from an inspection of evolutionary tracks:
\begin{align}
    & \frac{L_{\mathrm{low}}}{\Lsun} = \mathrm{max}\left(10^{2.5}, L_0(M)\right) \label{eq:LowLuminosityLimit} \\
    & L_0(M) = 3882\frac{M}{\Msun}-9294 \,. \label{eq:LowLumFunction}
\end{align}

For computational reasons, the calculation of
pulsation models is sequential in luminosity:
the first model is computed with a luminosity given by Eq.~\ref{eq:LowLuminosityLimit},
then additional models are computed by increasing the
total luminosity by a step $\Delta\log(L/\Lsun)=0.01$,
thus constructing a series of models having
the same mass, chemical composition and input
physics, but differing in luminosity and core mass (see Sect.~\ref{sec:CoreMassRadius}).
Such a sequence is terminated at a fixed
upper luminosity limit, $L=10^5\,\Lsun$, or
at the luminosity where convergence failure arises.
In either case, the resulting coverage is
compatible with that expected from
evolutionary calculations (Fig.~\ref{fig:MassLuminosityCoverage}).
We stress that such luminosity sequences,
used in the following sections to
examine the development of variability
in AGB models, are not evolutionary
tracks, and should not be regarded as such.

\subsubsection{Core mass and radius}
\label{sec:CoreMassRadius}

\begin{figure}
    \includegraphics[width=\columnwidth]{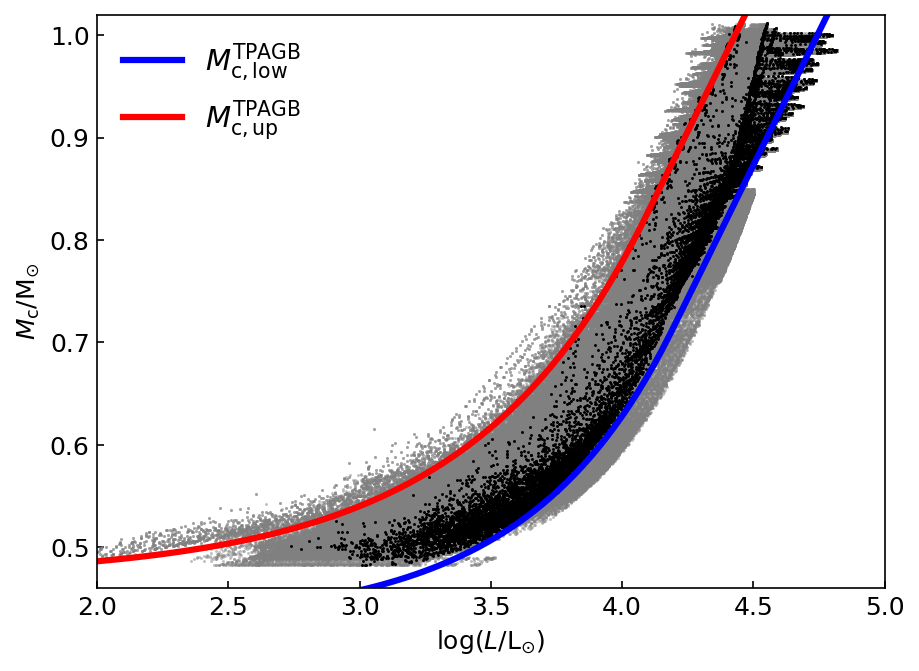}
    \caption{TP-AGB evolutionary models, computed with the \texttt{COLIBRI} code,
    in the core mass-luminosity plane. Grey dots in the background represent
    models including the full thermal pulse cycle, whereas black dots are
    quiescent phases. Solid lines represent the analytic relations
    for the upper (red) and lower (blue) limits of the region
    occupied by models.}
    \label{fig:CoreMassLuminosityPlane}
\end{figure}

It is well known that the total luminosity
of AGB stars is largely determined by the core mass,
a property usually known as `core mass-luminosity relation' (CMLR).
However, the classical CMLR breaks down in a number of
cases, most notably during the luminosity dip associated
with a He-shell flash, or in stars undergoing Hot-Bottom Burning (HBB).
As a result, evolutionary models do not follow
a tight relation in the $\Mc-\log(L)$ plane,
but are rather distributed in a broad stripe (Fig.~\ref{fig:CoreMassLuminosityPlane}).
To account for this, instead of limiting the
computation of envelope models to a single
value of core mass (\ie, a single CMLR),
we sampled the core mass-luminosity plane
and derived two analytic CMLR representative, respectively,
of the lower and upper limit values of core mass.
To do so, we made use of \texttt{COLIBRI}
TP-AGB evolutionary tracks \citep{Marigo_etal_2017},
and found that both the lower limit ($M_{\mathrm{c,low}}^{\tiny \mathrm{TPAGB}}(L)$)
and upper limit ($M_{\mathrm{c,up}}^{\tiny \mathrm{TPAGB}}(L)$) relations
are well described by a function of $L$ that is
linear at low luminosities and logarithmic at large luminosities.
The adopted functional form is given by
\begin{equation}\label{eq:CoreMassLuminosityFunctions}
    \frac{\Mc}{\Msun} = \left\{
    \begin{array}{lr}
    a_1 + a_2 \cdot 10^{a_3 \cdot \left[\log(L/\Lsun) + a_4\right]} & \mbox{if }\log(L/\Lsun) < \ell \\
    b_1 + b_2 \cdot \left[\log(L/\Lsun) + b_3\right] & \mbox{if }\log(L/\Lsun) \geq \ell
    \end{array}\right. \,,
\end{equation}
with coefficients given in Table~\ref{tab:McupMclowCoefficients}.

For each combination of the other grid parameters,
we computed two envelope models, one for each
of the two CMLRs. This approach represents
a significant improvement with respect to
the use of a single CMLR, in that the pulsation
properties can be estimated for arbitrary values
of core mass (for instance by interpolating in the two $\Mc$ values provided),
thus allowing a more accurate coverage of
the properties of TP-AGB stars.
To test this approach, we have computed envelope models
with values of core mass and luminosity that sample the
whole region between the two CMLRs defined above,
and extending substantially below the `lower limit' CMLR.
We examined the linear stability for those envelope
models and found that periods $\log(P_n)$ and
growth rates $GR$ depend weakly on the core mass,
and that the dependence is very close to linear.
The interpolation and extrapolation
of periods and growth rates as a function of core
mass can thus be considered safe.

The core radius, on the other hand,
has been kept fixed to a constant value $\Rc=0.15\,\Rsun$,
that is a bit larger than the radius of the core
(and of the H-burning shell) during
most of the TP-AGB evolution.
This choice implies the rather strong
assumption that pulsation properties
are at most weakly dependent upon the core radius.
We verified that if the radius is instead
obtained from a mass-radius relation for
white dwarfs (physically more appropriate),
there is no appreciable difference in the
resulting periods and growth rates.

\begin{table}
\normalsize
\centering
\caption{
    Coefficients of the functions describing the upper ($M_{\mathrm{c,up}}^{\mathrm{TPAGB}}$)
    and lower ($M_{\mathrm{c,low}}^{\mathrm{TPAGB}}$)
    boundaries of the region populated by TP-AGB evolutionary
    models in the $\Mc$--$\log(L)$ plane. See also Fig.~\ref{fig:CoreMassLuminosityPlane}.
    }
\label{tab:McupMclowCoefficients}
\begin{tabular}{c|c|c}
%\hline
& $M_{\mathrm{c,low}}^{\mathrm{TPAGB}}$ & $M_{\mathrm{c,up}}^{\mathrm{TPAGB}}$ \\
\hline
\hline
$\ell$  & $4.19$                & $4.08$ \\
$a_1$           & $0.425$               & $0.4708$ \\
$a_2$           & $9.425\cdot10^{-6}$   & $2.293\cdot10^{-3}$ \\
$a_3$           & $0.7884$              & $0.6468$ \\
$a_4$           & $1.4964$              & $-0.7103$ \\
$b_1$           & $-1.52$               & $-1.279$ \\
$b_2$           & $0.5202$              & $0.5202$ \\
$b_3$           & $0.1$                 & $-0.05$ \\
\end{tabular}
\end{table}

\subsubsection{Effective temperature}
\label{sec:EffectiveTemperature}

The effective temperature is an output
parameter from the integration of a static envelope model,
and is essentially determined by the choice
of the mixing length parameter $\aml$.
In evolutionary models, the mixing
length parameter is calibrated by reproducing the
observed properties of the present Sun.
Envelope models such as the one used
here are not coupled to any evolutionary
calculation, so this procedure is not possible.
Instead, $\aml$ is usually tuned
to reproduce some reference value
of the effective temperature.
For instance, in their modelling of
LPVs in globular clusters,
\citet{Lebzelter_Wood_2007} and \citet{Lebzelter_Wood_2016}
calibrated the mixing length parameter
to reproduce the slope of the observed giant branch,
obtaining values of $\aml$ in the range $\sim1.7$ to $\sim2.2$.
When computing envelope models with the requirement
of reproducing the effective temperatures
of \texttt{COLIBRI} models, we found a
range of $\aml$ consistent with those values
\footnote{
    Differences in $\Teff$ between envelope
    models and evolutionary models when
    using the same value of $\aml$ are
    primarily due to the different
    input physics
    and boundary conditions
    employed in the two codes.
}.

Therefore, we computed each pulsation
model with three different values of
the mixing length parameter: $\aml=1.5$, $2.0$, $2.5$.
As a result, pulsation models in the
grid, for a given combination of mass,
core mass, luminosity and chemical composition,
span a finite range of effective temperatures
(using $\aml=2.5$ will result in a $\sim$15--20 per cent
change in effective temperature, depending on other input parameters
with respect to the case in which $\aml=1.5$ is used).
The major advantage of this is that
pulsation models presented here can be used to estimate
pulsation properties from the global
parameters of any stellar model,
regardless of the code it originates
from and the corresponding calibration of $\aml$.

\subsection{Pulsation models}
\label{sec:PulsationModels}

For each envelope model in the grid
we computed pulsation properties
by performing linear stability analysis
with the code described in \citet{Wood_Olivier_2014}.
As for the envelope models, convection
is described by the mixing length theory \citep{Cox_Giuli_PSS_1968}.
The interaction between convection and pulsation
is treated in the simplified manner
described in \citet{Fox_Wood_1982} based
on the formalism by \citet{Arnett_1969},
while models do not include turbulent pressure or the
kinetic energy of turbulent motions.
Turbulent viscosity, described in the pulsation
code in terms of a free parameter $\alpha_{\nu}$
\citep[see][for more details]{Wood_Olivier_2014,Keller_Wood_2006},
is sometimes included in non-linear calculations
to bring predicted pulsation amplitudes in
agreement with observations \citep[see, \eg,][]{Ireland_etal_2008},
and contributes to the damping of oscillations in linear pulsation models.
However, since the turbulent viscosity parameter
lacks a proper calibration as a function
of global stellar parameters, we set $\alpha_{\nu}=0$ in all models.

The result of the stability analysis applied to an envelope model
consists of a set of complex eigenfrequencies.
The number of pulsation modes to be searched is user-defined,
and in the present work we computed five modes for each model.
Modes are identified by their radial order $n$, with $n=0$
corresponding to the fundamental mode, $n=1$ to the first overtone
(1O) mode, $n=2$ to the second overtone (2O) mode, and so on.
The time dependence of the perturbations is assumed to be in the
form $\xi\propto\exp(\omega t)$, with eigenfrequency $\omega=\omega_{\rm R} + \mathbf{i}\omega_{\rm I}$.
The imaginary part of $\omega$ is the angular frequency of oscillation,
and the period of the $n$-th mode is defined as:
\begin{equation}\label{eq:P_def}
    P_n = \frac{2\pi}{\omega_{{\rm I},n}} \,.
\end{equation}
While it is generally accepted that all red giants are variable to some extent
\citep[\eg,][and references therein]{Percy_etal_2001},
not all of them exhibit photometric variations to a detectable level.
Whether or not an oscillating star can be observed as variable depends on the
photometric amplitude of its variability, as well as on the instrumental setup.
In multi-periodic variables, the amplitude of a given mode with
respect to the others is also important. In fact, LPVs are often
characterised in terms of the observed properties of
the dominant (or primary) mode, the one with the largest amplitude.

Since linear models are not able to predict pulsation amplitudes, we rely
on a different approach to allow for the comparison with observations.
The amplitude of a pulsation mode depends upon its degree of excitation,
which is described by the non-adiabatic term of the eigenfrequency,
\ie, the real part $\omega_{\rm R}$ (also known as stability coefficient).
It is proportional to the inverse timescale of exponential decay/amplification
of the perturbation. A mode with $\omega_{\rm R}<0$ is stable, and hence we expect 
that it is not observable, whereas an unstable mode has $\omega_{\rm R}>0$,
and it should be observable provided it has had the time to grow to a large enough amplitude.
This latter condition corresponds to the requirement that $\omega_{\rm R}$
is larger than some (properly calibrated) threshold. 
In the following we will use the amplitude growth rate,
\begin{equation}\label{eq:GR_def}
    GR_n = \exp\left(2\pi\frac{\omega_{{\rm R},n}}{\omega_{{\rm I},n}}\right) - 1 \,,
\end{equation}
as a proxy for excitation. 
This quantity describes the fractional rate of change in the radial
amplitude per pulsation cycle \citep[see also][]{Fox_Wood_1982, Wood_Olivier_2014}.
This choice is justified by the findings of \citet{Trabucchi_etal_2017},
who used some of the models presented here to study the population of LPVs
in the Large Magellanic Cloud, and found that (1) growth rates are able
to account for the observed instability strip, and (2) observed
variability amplitudes scale with the predicted growth rates.
The relation between the growth rate and the stability
coefficient is further discussed in Sect.~\ref{sec:StabilityOvertones}.
We will assume hereafter that the dominant pulsation mode in a given
envelope model corresponds to the one with the largest growth rate.
It is important to keep in mind that, while the results of \citet{Trabucchi_etal_2017}
are encouraging in this respect, growth rates depend on the treatment
of poorly understood processes such as convection and its coupling
with pulsation, and are therefore affected by uncertainties.

A significant fraction of the LPVs in the Milky Way and in the Magellanic Clouds
with $I$-band magnitude fainter than the tip of the Red Giant Branch (TRGB)
have been recognised as RGB stars
\citep[see, \eg,][]{Ita_etal_2002, Ita_etal_2004, Kiss_Bedding_2003,Soszynski_etal_2004_OSARG}.
As explained in previous sections,
some of the parameter-domains of the present grid
have been defined using as reference TP-AGB models. 
We examine here whether the grid coverage
is also appropriate for the global parameters
of RGB stars down to $\sim1$ mag below the TRGB.
Based on OGLE-III data \citep{Soszynski_etal_2009_LMC},
that limit would correspond to $\ks\simeq14$ mag,
or $\log(L/\Lsun)\gtrsim2.5$, meaning that our grid coverage
is sufficient to describe OSARGs in the RGB phase.

At a given luminosity, the main differences
between RGB and AGB stars in terms of global
parameters are effective temperature,
mass and core mass. The mass coverage in the grid
($0.6\leq M/\Msun\leq7.0$) is safely appropriate for
RGB stars in the regime of interest for long-period variability,
and the three nodes of $\aml$ used in constructing the
grid cover a wide enough range of temperatures to
account for the slightly colder RGB stars.
On the other hand, the core mass in RGB stars is smaller with respect
to AGBs at the same luminosity, and actually falls
below the `lower limit' CMLR discussed in Sect.~\ref{sec:CoreMassRadius}.
In other words, the use of our models to predict pulsation properties
of RGB stars, \eg, by interpolation in the grid
(see Appendix~\ref{sec:InterpolationRoutine})
will result in extrapolation below the lower $\Mc$ boundary of the grid.
However, for low $\Mc/M$ ratios, predicted periods
and growth rates of radial modes appear to depend
weakly on the core mass. Moreover, the dependence
is very close to linear, suggesting that even
when extrapolating towards lower masses one
would obtain values close to those resulting
from direct calculations (see Sect.~\ref{sec:CoreMassRadius}).
According to these arguments, the grid coverage is
appropriate for the global parameters
of RGB-OSARGs.

Finally, it should be kept in mind that
these considerations are only valid
under the assumption that the core is
dynamically decoupled from the envelope,
an approximation that is generally valid also for luminous RGB stars.

\begin{table*}
\footnotesize
\centering
\caption{
    Example of results for a selected series of pulsation models.
    Defining parameters of the sequence are shown in the header.
    The complete set of tables is available for download
    from the website \url{http://starkey.astro.unipd.it/web/guest/pulsation} (see Appendix~\ref{sec:WebInterface}).
    }
\label{tab:example_table}
\begin{tabular}{ccc|ccc|ccc|ccc}
\multicolumn{12}{c}{$M=1.5\,\Msun \,,\; \aml=2.0 \,,\; {\rm CMLR:}\;M_{\mathrm{c,low}}^{\mathrm{TPAGB}}$} \\
\multicolumn{12}{c}{$\Zref=0.006 \,,\; X=0.7 \,,\; \fco=0. \,,\; \co=(\co)_{\odot}=0.54954$} \\
\hline
\hline
$\log$    & $\log$  &             & \multicolumn{9}{c}{(angular frequencies in $\mu$Hz)} \\
$L/\Lsun$ & $\Teff/\mathrm{K}$ & $\Mc/\Msun$ & $\omega_{\rm ad,0}$ & $\omega_{\rm R,0}$ & $\omega_{\rm I,0}$ & $\omega_{\rm ad,1}$ & $\omega_{\rm R,1}$ & $\omega_{\rm I,1}$ & $\omega_{\rm ad,2}$ & $\omega_{\rm R,2}$ & $\omega_{\rm I,2}$ \\
\hline
\multicolumn{3}{c|}{...} & \multicolumn{3}{c|}{...} & \multicolumn{3}{c|}{...} & \multicolumn{3}{c}{...} \\
 3.70 & 3.551 & 0.543 & 0.4607 & -0.0011 & 0.4628 & 0.8402 & 0.0143 & 0.8624 & 1.2944 & -0.0082 & 1.3439 \\
 3.71 & 3.550 & 0.545 & 0.4453 & -0.0011 & 0.4474 & 0.8176 & 0.0142 & 0.8396 & 1.2610 & -0.0088 & 1.3110 \\
 3.72 & 3.549 & 0.547 & 0.4304 & -0.0012 & 0.4323 & 0.7956 & 0.0142 & 0.8174 & 1.2284 & -0.0093 & 1.2788 \\
 3.73 & 3.548 & 0.549 & 0.4159 & -0.0012 & 0.4177 & 0.7742 & 0.0142 & 0.7958 & 1.1966 & -0.0099 & 1.2476 \\
 3.74 & 3.546 & 0.552 & 0.4018 & -0.0012 & 0.4035 & 0.7534 & 0.0142 & 0.7747 & 1.1656 & -0.0104 & 1.2169 \\
 3.75 & 3.545 & 0.554 & 0.3880 & -0.0012 & 0.3896 & 0.7331 & 0.0142 & 0.7543 & 1.1354 & -0.0108 & 1.1871 \\
 3.76 & 3.544 & 0.556 & 0.3747 & -0.0012 & 0.3761 & 0.7133 & 0.0142 & 0.7343 & 1.1058 & -0.0113 & 1.1580 \\
 3.77 & 3.543 & 0.558 & 0.3616 & -0.0012 & 0.3629 & 0.6939 & 0.0142 & 0.7146 & 1.0766 & -0.0117 & 1.1293 \\
 3.78 & 3.542 & 0.561 & 0.3490 & -0.0012 & 0.3502 & 0.6752 & 0.0142 & 0.6957 & 1.0485 & -0.0123 & 1.1016 \\
 3.79 & 3.540 & 0.563 & 0.3368 & -0.0011 & 0.3378 & 0.6570 & 0.0142 & 0.6773 & 1.0211 & -0.0126 & 1.0747 \\
 3.80 & 3.539 & 0.566 & 0.3249 & -0.0011 & 0.3257 & 0.6393 & 0.0142 & 0.6594 & 0.9943 & -0.0130 & 1.0485 \\
\multicolumn{3}{c|}{...} & \multicolumn{3}{c|}{...} & \multicolumn{3}{c|}{...} & \multicolumn{3}{c}{...} \\
\end{tabular}
\end{table*}

Note also that \citet{Trabucchi_etal_2017} used
some of the models presented here to study both the AGB and the RGB
populations of LPVs in the Large Magellanic Cloud,
obtaining reasonably good agreement with observations.
We therefore consider our models suitable enough
to study radial oscillations in bright RGB stars,
provided the arguments above are taken into account
in the interpretation of the results.

The results of the models are collected in tables
giving global stellar parameters, together
with the adiabatic frequency and the real and
imaginary parts of the non-adiabatic
frequency for five radial modes, from the
fundamental to the 4O mode. Table~\ref{tab:example_table}
shows a typical output as an example,
displaying frequencies for the fundamental, 1O, and 2O modes.

%%%%%%%%%%%%%%%%%%%%%%%%%%%%%%%%%%%%%%%%%%%%%%%%%%%%%%%%%%%%%%%% STABILITY

\section{Pulsational Stability}
\label{sec:Stability}

In the present section we examine the stability
of pulsation in different radial modes, and how
it depends upon global stellar parameters.
To begin with, we consider a series of models
of increasing radius and luminosity, and with
fixed mass, composition, and input physics,
\ie, one of the luminosity sequences of the grid.
Values of the fixed quantities are summarised
in Table~\ref{tab:SelectedSequence}, and are
representative of typical O-rich TP-AGB variables in the LMC.
Changing the other parameters does not alter
significantly the picture provided here,
except in the high-mass range of the grid.
This is discussed in Sect.~\ref{sec:DependenceGrowthRatesStellarParameters}.

The behaviour of growth rates as a function
of increasing luminosity allows us to present
a general picture of the evolution
of pulsational stability during the AGB.
Compared to full evolutionary tracks,
it provides us with a simplified framework, suitable
for understanding the main factors determining
the global evolution of LPVs. We complete this
picture in Sect.~\ref{sec:PulsationEvoTrack},
where periods and growth rates along proper
evolutionary tracks are analysed.

\begin{figure}
    \includegraphics[width=\columnwidth]{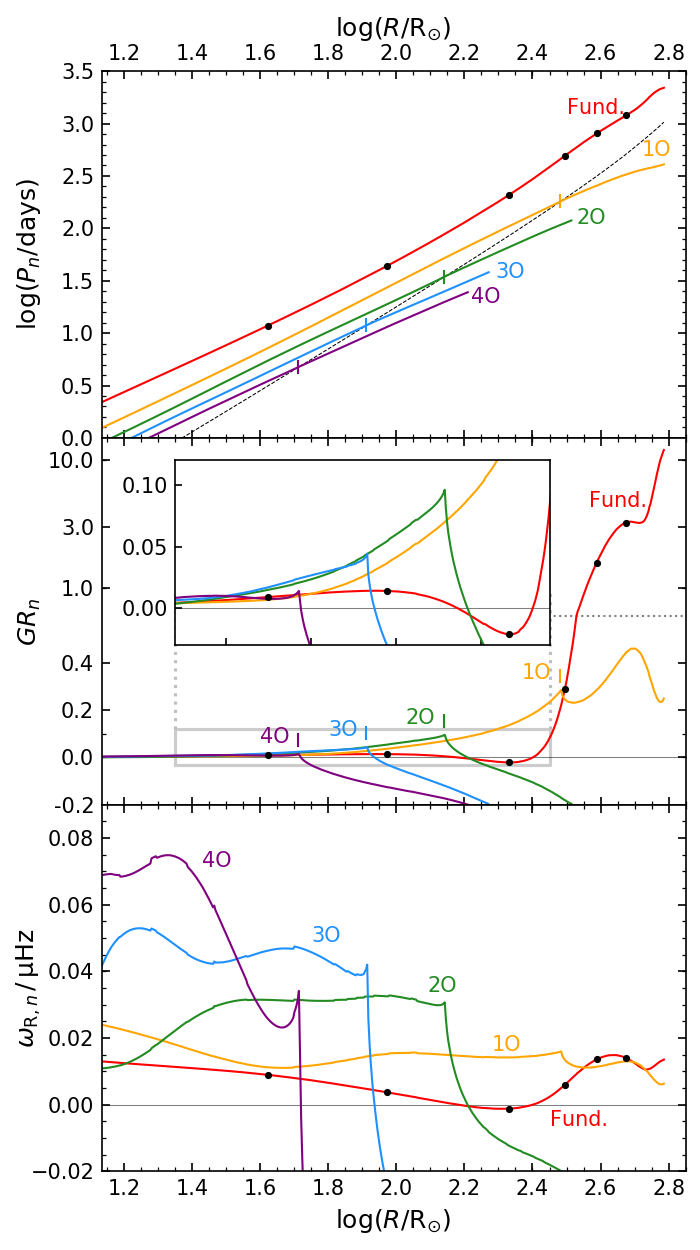}
    \caption{Global pulsation properties as a function of stellar radius for a selected
    series of models (see Table~\ref{tab:SelectedSequence}).
    Top panel: pulsation periods (solid lines) and the period corresponding to
    the acoustic cut-off (see Sect.~\ref{sec:StabilityOvertones}).
    Middle panel: amplitude growth rates (Eq.~\ref{eq:GR_def});
    the inset shows an exploded view around $GR=0$.
    Note that the vertical scale is linear below $GR=0.6$,
    but becomes logarithmic above this value (dotted horizontal line)
    to better illustrate the rise of $GR$ for the fundamental mode.
    Bottom panel: stability coefficients $\omega_{{\rm R},n}$.
    Vertical ticks mark the peaks of overtone modes
    growth rates in the top and middle panels. Black filled circles
    correspond to the selected models shown in Fig.~\ref{fig:SelectedFundMode_W_G1}.
    }
    \label{fig:SelectedSequence}
\end{figure}

Fig.~\ref{fig:SelectedSequence} shows periods and growth rates for the
five lowest-order radial modes as a function of radius
along the selected luminosity sequence. Given the dynamical
nature of stellar pulsation, periods scale approximately
as the free-fall time of the star (proportional to
the inverse square root of the mean stellar density),
and thus increase with radius approximately as a power-law.

Growth rates $GR$ show a more complex, characteristic pattern.
At low radii, all modes have rather small
(or even negative) growth rates, meaning that
stars at this stage would exhibit only low-amplitude
variability, or no detectable variability at all\footnote{
    While the present models are limited to five radial
    modes, calculations including a larger number of
    modes have been carried out, and we did not find
    any mode to deviate from the picture given here.
}. Growth rates increase with luminosity, following a trend
that is qualitatively the same for all modes, with the exception
of the fundamental mode. Overtone modes growth rates increase
more or less steadily until they reach a peak-shaped
maximum, after which they undergo a quick drop
towards negative values. Note that after this stage the growth rate
never recovers to positive values: once an overtone
mode has passed through its stage of maximum growth rate, it becomes
definitively stable. In a real AGB star, this is true as long as the
stellar radius grows with time: if, on the other
hand, the star shrinks, \eg, after a thermal pulse,
a stable overtone mode can become unstable again
(see Sect.~\ref{sec:PulsationEvoTrack}).
The 1O mode represents an exception
to this, as its growth rate usually keeps positive values
even past the `peak': in principle, it should be
detectable with finite amplitude up to the highest luminosities.

Beginning from the low-luminosity end of a sequence,
the highest radial order in our calculations, the 4O mode,
is the first to reach the maximum, followed by
the 3O, 2O and 1O modes in this order.
When a mode is approaching its point of maximum instability,
its growth rate is larger than that of all other modes,
and we tag it as `dominant' (see the
discussion in Sect.~\ref{sec:StabilityFit}).
In terms of the evolution of a LPV, this means that at
low luminosity the dominant mode is a relatively
high-order overtone \citep[probably the 3O mode, see][]{Trabucchi_etal_2017}.
As the star evolves, that mode becomes stable, and
the overtone of immediately lower order takes over
as dominant. This event is repeated in a series
of `shifts' towards lower order modes, until
eventually the fundamental mode becomes dominant,
a well known characteristic of red giant variables
(see, \eg, \citealt{Lattanzio_Wood_2003}, and \citealt{Xiong_etal_2018} and references therein).

The behaviour of the fundamental mode growth rate is
clearly different, and is discussed in Sect.~\ref{sec:StabilityFundamental}.

\label{sec:StabilityOvertones}

\begin{table}
\normalsize
\centering
\caption{
    Global parameters of the selected series of models
    discussed in Sect.~\ref{sec:Stability} and \ref{sec:Periods},
    and displayed in Fig.~\ref{fig:SelectedSequence}
    }
\label{tab:SelectedSequence}
\begin{tabular}{l|l}
Parameter       &   Value                       \\
\hline
$M/\Msun$       &   1.5                         \\
$Z\,(=\Zref)$   &   0.006                       \\
$X$             &   0.7                         \\
$\co$           &   $(\co)_{\odot}\simeq0.55$   \\
$\aml$          &   2.0                         \\
$\Mc$           &   `lower limit' CMLR        \\
\end{tabular}
\end{table}

\begin{figure}
    \includegraphics[width=\columnwidth]{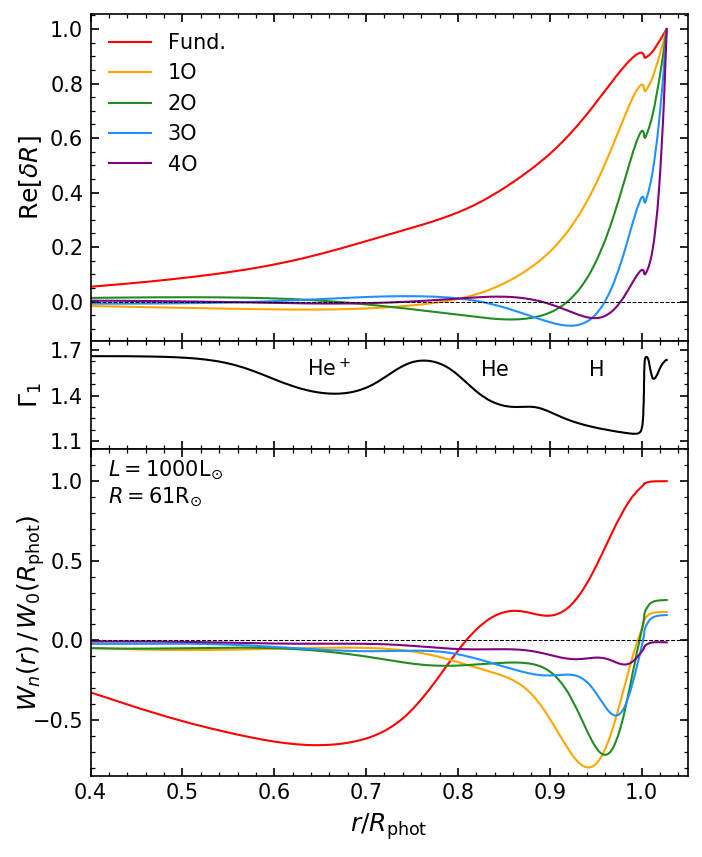}
    \caption{Internal pulsation properties for the model with
    $L=1000\,\Lsun$, $R=61\,\Rsun$ and belonging to the selected luminosity sequence
    (see Table~\ref{tab:SelectedSequence}).
    Radial displacement eigenfunctions and partial work integrals as
    a function of fractional radius are displayed in the top and bottom
    panels, respectively. To simplify the visual comparison, the partial work integrals are normalised
    to the surface value of the fundamental mode. The adiabatic exponent
    $\Gamma_1$ is shown in the middle panel, highlighting the
    regions of partial ionization and molecular dissociation of major elements. The interior
    amplitude of overtone modes is essentially zero everywhere except
    for the outer $\lesssim 10$ per cent of the stellar radius, corresponding
    to the outer part of the H ionization zone. The 1O mode extends
    a bit deeper (but still with low interior amplitude) down to
    the first He ionization zone. The fundamental mode is the only one
    with significant interior amplitude in a large portion of the envelope,
    including all major ionization zones. Regions above $\sim65$ per cent
    of the radius contribute to driving the fundamental mode,
    while overtone driving regions are limited to the outer part
    of the H ionization zone.}
    \label{fig:SelectedEFW}
\end{figure}

Overtone modes are essentially confined near
the surface of the models, as displayed in
the top panel of Fig.~\ref{fig:SelectedEFW}.
Their properties are thus in large part
determined by physical conditions in the outermost
stellar layers. We make use of the partial work
integral $W(r)$ \citep[see, \eg,][]{Cox_TSP_1980, Fox_Wood_1982}
to understand the driving of pulsation.
$W(r)$ represents the amount of work done per unit time
by all mass layers interior to a certain position $r$.
Driving regions, where $W(r)$ increases outward,
provide positive work, \ie, feed energy into pulsation motion.
All the overtone modes have a significant amount of driving coming from the outer
part of the partial hydrogen ionization zone.
To examine what happens as models evolve, for each mode
we consider both the growth rate and the
stability coefficient $\omega_{{\rm R},n}$
the latter being proportional to the work integral.
A clear difference between $GR$ and $\omega_{\rm R}$,
evident in Fig.~\ref{fig:SelectedSequence},
is that the former grows with increasing radius,
while the latter is approximately constant.
This means that the increasing instability
(and amplitude) experienced by overtone modes
is not due to an increase in the rate
of work done, but rather to the fact that
the amount of work is integrated over
a longer pulsation period (compare with the definition of growth rate, Eq.~\ref{eq:GR_def}).

This is true, at least, prior to the point of
maximum growth rate, after which $\omega_{\rm R}$
also drops to negative values. However, this
abrupt stabilisation is not due to driving processes:
it is a purely mechanical effect determined
by the choice of realistic boundary conditions in the
pulsation models (described in \citealt{Wood_Olivier_2014}).
The propagation of acoustic perturbations in the form
of standing waves (\ie, pulsation modes) in a stellar envelope
occurs only if their frequency is smaller than a critical
value, the acoustic cut-off frequency $\nuac$,
determined by the surface properties of envelope models.
When a mode's frequency exceeds $\nuac$,
pressure waves travelling towards the surface are not
reflected back to the interior, but will rather
propagate outward through the atmosphere,
dissipating the pulsation energy and thus rendering
the mode stable.

The regular stabilisation pattern of overtone
modes can be easily understood if we consider
the expression of the acoustic cut-off frequency
in the approximation of an isothermal atmosphere:
\begin{equation}\label{eq:nuac_isothermal}
    \nuac = \frac{1}{4\pi}G\left(\frac{m_{\rm H}}{k_{\rm B}} \right)^{1/2}\left(\frac{\mu\Gamma_1}{T}\right)^{1/2}\frac{M}{R^2} \,,
\end{equation}
where $G$ is the gravitational constant, $m_{\rm H}$ is the atomic mass unit,
$k_{\rm B}$ is the Boltzmann constant, $\mu$ is the mean molecular weight,
$\Gamma_1=({\rm d}\ln P/{\rm d}\ln\rho)_{\rm ad}$ is the adiabatic exponent,
$M$ and $R$ are the surface mass and radius, and $T\sim\Teff$ is the surface temperature.
While $\nuac\propto M R^{-2}$, the oscillation frequency of
a given overtone mode of radial order $n$
scales as $\nu_n=P_n^{-1}\sim M^{1/2}R^{-3/2}$. This means that both frequencies
decrease as the star expands, but $\nuac$ does so at a faster rate,
and eventually the two cross each other, as displayed in Fig.~\ref{fig:SelectedSequence}
\citep[see also fig.~3, top panel, of][]{Wood_Olivier_2014}.
Maxima of the growth rates occur exactly when $\nu_n=\nuac$.

The abrupt stabilisation of
overtone modes that exceed
the acoustic cut-off is the very
reason for the right edges of the PL sequences of
LPVs observed in the Magellanic Clouds
(with the exception of sequence C
associated with fundamental mode pulsation).
As a consequence of this behaviour, overtone
modes are dominant, and most easily observable,
when their frequency is close to the acoustic cut-off value.

Remarkably, this behaviour is similar to that
followed by solar-like oscillations in less luminous red giants.
Solar-like oscillations are observed in main sequence and post-main stars
with a convective envelope and $\Teff\lesssim6500K$.
Their oscillation spectra present radial and non-radial modes that are
stochastically excited and damped by sub-photospheric convection.

Ultra-precise photometric observations by the space telescopes 
CoRoT \citep{Baglin_etal_2009} and \textit{Kepler} \citep{Borucki_etal_2009, Borucki_etal_2010}
have allowed the characterisation of the properties of solar-like spectra
for evolutionary stages going from the main-sequence to the evolved red giant region.
The global properties of these spectra are defined by two parameters,
$\numax$ and $\langle\Delta\nu\rangle$, that depend on global stellar
properties \citep[see \eg][]{Chaplin_miglio_2013}.  
The power spectrum consists in a set of regularly spaced
radial and non-radial modes whose amplitudes appear
modulated by a gaussian-shape envelope centred
at $\numax$ and FWHM (for red giants) given by
$\sigma\simeq 0.28\,\numax^{0.15}\,\langle\Delta\nu\rangle$ (\citealt{Mosser10}).
The frequency at maximum power, $\numax$,
appears to scale with the acoustic cut-off frequency
(see \citealt{Brown_etal_1991,Kjeldsen_Bedding_1995, Belkacem_etal_2011},
and the recent improvement by \citealt{Viani_etal_2017}
that includes also the dependence on the mean molecular weight).
The so-called large frequency separation, $\langle\Delta\nu\rangle$,
is the difference in frequency between modes with consecutive radial order and same angular degree,
and, in the asymptotic regime, scales with $\bar\rho^{1/2}$
(the square root of stellar mean density, \citealt{Ulrich_1986, Kjeldsen_Bedding_1995}).
Taking into account the asymptotic expression for
the frequency of radial modes ($\nu_{n}=(n+\epsilon)\langle\Delta\nu\rangle$
it is obvious that as a star of mass $M$ evolves along the RGB the radial order
of the mode corresponding to the maximum power ($n_{\rm max}$)
shifts to lower values as radius increases ($n_{\rm max} \propto (M/(R\,\Teff))^{1/2}$).
Moreover, the mean density is decreasing, and so is the frequency domain
of the oscillation spectrum, leading to a smaller number of radial orders in the spectrum.
The kind of `universal pattern' shown by oscillations in red giants
allowed \cite{Mosser_etal_2013} to link the most luminous variables
observed by {\textit Kepler} with LPVs in OGLE.

Low-order overtone modes in LPVs and high-luminosity
solar-like pulsators on the red giant branch share the property of showing
the largest amplitude for frequencies approaching to the cut-off value.
Whether or not the excitation mechanism is the same
for both kinds of pulsators is still matter of debate
\citep[\eg,][]{JCD_etal_2001, Dziembowski_etal_2001, Xiong_Deng_2013},
and the fact above described should not be immediately interpreted
as an evidence for a common excitation mechanism,
but rather as an indication that properties
of low-order radial modes are largely determined
by stellar global properties (surface gravity and temperature).

\subsection{Stability Evolution of the Fundamental Mode}
\label{sec:StabilityFundamental}

\begin{figure}
    \includegraphics[width=\columnwidth]{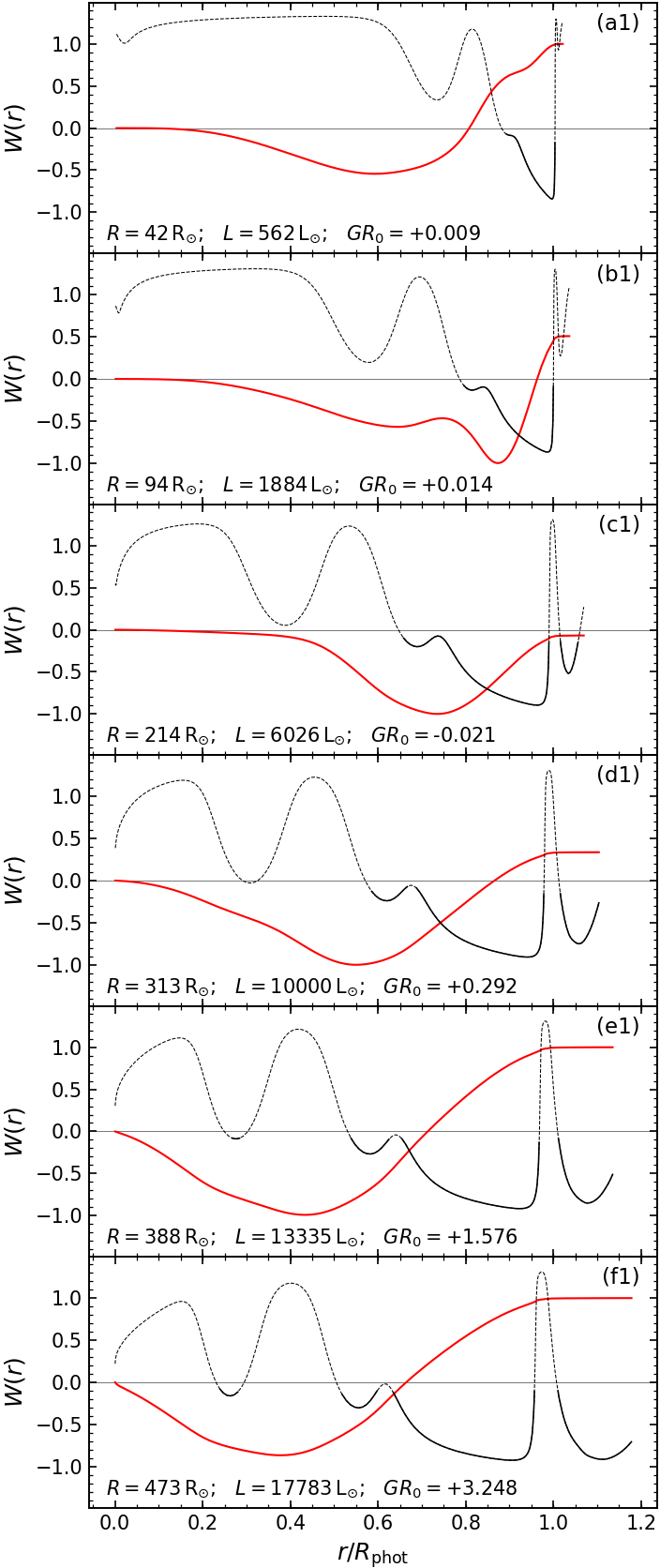}
    \caption{Partial work integral (normalised to the absolute maximum)
    of the fundamental mode for six models in the
    selected luminosity sequence (Table~\ref{tab:SelectedSequence}).
    The radial profile of the adiabatic exponent $\Gamma_1$ is also
    shown to highlight the location of partial ionization and molecular dissociation zones of major elements
    (dashed line, shown as solid line instead where $\Gamma_1 < 4/3$).
    Models shown here are marked by black filled circles in Fig.~\ref{fig:SelectedSequence}
    and Fig~\ref{fig:Selected_dPdR}. Luminosity increases from top to bottom panels.}
    \label{fig:SelectedFundMode_W_G1}
\end{figure}

The evolution of the fundamental mode growth rate
is intrinsically different from that of the overtone modes.
It does not increase monotonically with radius, but rather shows an
intermediate phase of decrease during which
it reaches a minimum value (depending on mass,
see Sect.~\ref{sec:DependenceGrowthRatesStellarParameters}).
In models of intermediate mass ($1\lesssim M/\Msun \lesssim 4$,
depending on chemistry and input physics)
the fundamental mode growth rate is negative at the minimum,
implying stability. This stage is followed by a
very steep increase, causing the linear growth rates
of the fundamental mode to become extremely large ($GR_0\gg1$)
at high luminosity. This would mean that pulsation
amplitudes should increase by a factor of several
at each pulsation cycle. This is clearly unrealistic from a physical
point of view, and stems from the
fact that the linear approximation for oscillations
tends to break down at high luminosities.
Nonetheless, this result is qualitatively meaningful,
and we interpret it as a phase of large-amplitude
fundamental mode pulsation (in other words, the Mira-like phase).

In contrast with overtone modes, the (temporary)
stabilisation of the fundamental mode is not
caused by its frequency reaching the acoustic cut-off.
In fact, the present models suggest that the fundamental
mode never reaches the cut-off (Fig.~\ref{fig:SelectedSequence}, top panel).
Moreover, the fundamental mode would reach the cut-off frequency
at larger radii than the 1O mode, while the
fundamental mode stabilisation actually precedes
the peak of the 1O mode's growth rate.
It is thus clear that the fundamental mode growth rate is affected
by actual non-adiabatic processes taking place within the envelope.
A detailed investigation of the driving mechanism
of the fundamental mode, with particular attention
to the role of convection, is highly desirable
to properly interpret the results of pulsation models.
This kind of analysis, however, is beyond the scope
of the present work, and will be the subject of
a forthcoming study. We limit the discussion here
to the observation that the growth rate of the
fundamental mode is strongly sensitive to the
internal structure of the envelope, and in particular
to the displacement of the region of ionization
of major elements. This is displayed in Fig.~\ref{fig:SelectedFundMode_W_G1},
where the partial work integral $W(r)$ (solid red line) of the fundamental
mode is shown for a few models
along the selected luminosity sequence
(marked by black filled circles in Fig.~\ref{fig:SelectedSequence}).
It is worth pointing out that this property
of the fundamental mode makes it difficult
to relate either its growth rate or its
period to global stellar parameters in a
simple and straightforward way (see also Sect.~\ref{sec:PeriodsFit}).

\subsection{Dependence of growth rates upon stellar parameters}
\label{sec:DependenceGrowthRatesStellarParameters}

\begin{figure}
    \includegraphics[width=\columnwidth]{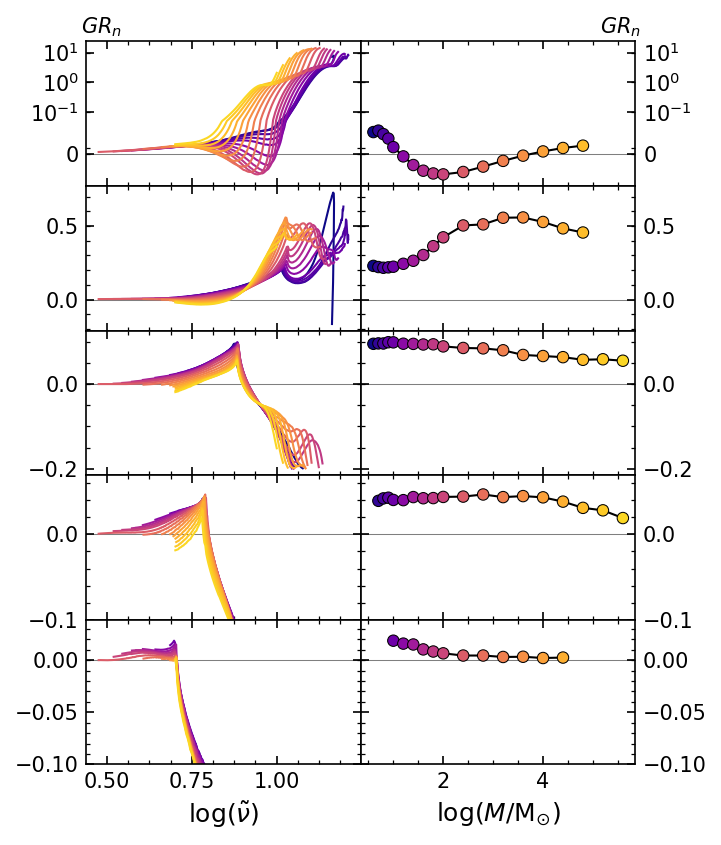}
    \caption{Left panels: growth rates as a function of $\tilde{\nu}$
    (see text) for modes from the fundamental (top) to the 4O (bottom).
    Each line represent a series of models with a given mass,
    colour coded as in the right panel.
    Global parameters other than mass are those summarised in Table~\ref{tab:SelectedSequence}.
    Right panels: value $GR_{\rm max,n}$ of the growth rate at
    the acoustic cut-off for each series of models, as a function
    of the corresponding mass. For the fundamental mode, the value
    of the growth rate at the minimum is shown instead.}
    \label{fig:GR_dep_M}
\end{figure}

The discussion in the preceding sections is limited
to the case of a specific mass, composition, and input physics.
Here, we expand the discussion to different input parameters.
Note that the total radius, used above as a proxy for
the evolutionary status, is not convenient to
compare the evolution of stability in models
with significantly different global parameters.
On the other hand, we have seen that the evolution
of growth rates is determined by the values
of the oscillation frequency with respect to
the acoustic cut-off frequency.
It is known from the theory of stellar
pulsation \citep[\eg][]{Cox_TSP_1980}
that oscillation frequencies scale as
$\nu\sim\bar\rho^{1/2}$, where
$\bar\rho=3M/4\pi R^3$ is the mean stellar density. Thus
it is convenient to define the quantity
\begin{equation}\label{eq:tildenu}
    \tilde{\nu} = \left(\frac{M}{\Msun}\right)^{-1/2} \left(\frac{R}{\Rsun}\right)^{1/2} \left(\frac{\Teff}{{\rm T}_{{\rm eff},\odot}}\right)^{1/2} \propto \frac{\nu}{\nuac} \,,
\end{equation}
to be used in place of the total radius in our analysis.

\begin{figure}
    \includegraphics[width=\columnwidth]{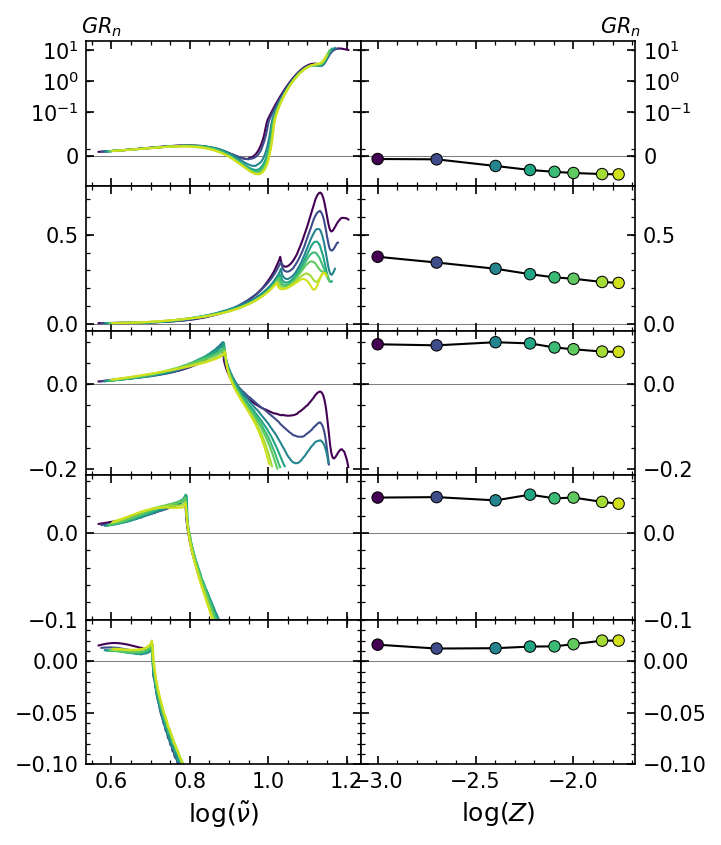}
    \caption{Same as Fig.~\ref{fig:GR_dep_M}, but showing series
    of models with different metallicity.
    Other global parameters are those summarised in Table~\ref{tab:SelectedSequence}.}
    \label{fig:GR_dep_Z}
\end{figure}

\begin{figure}
    \includegraphics[width=\columnwidth]{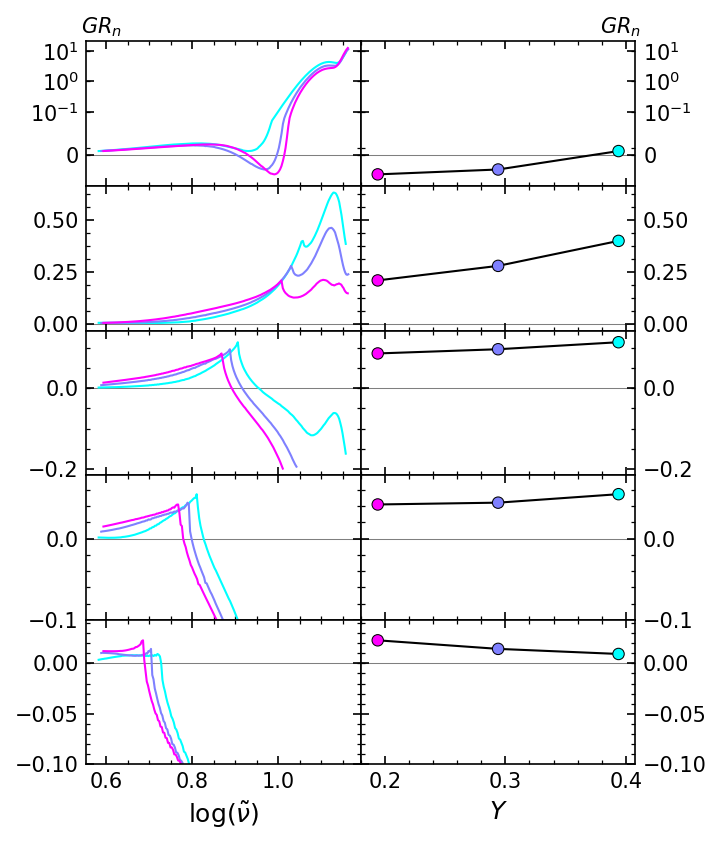}
    \caption{Same as Fig.~\ref{fig:GR_dep_M}, but showing series
    of models with different helium content.
    Other global parameters are those summarised in Table~\ref{tab:SelectedSequence}.}
    \label{fig:GR_dep_XY}
\end{figure}

\subsubsection{Varying mass}
The advantage of this newly defined parameter becomes
clear in Fig.~\ref{fig:GR_dep_M}, where growth rates are displayed
as a function of $\tilde{\nu}$ for series of models
with different masses. Note that $\tilde{\nu}$
increases as a star evolves to higher luminosities
even though $\nu_n$ decreases. This is because $\nu_n$
decreases with advancing evolution less rapidly than $\nuac$.
It is clear that in Fig.~\ref{fig:GR_dep_M}, for a given overtone,
growth rates reach the `peak' at approximately the same value
of $\tilde{\nu}$ for all masses, and the same is true for
varying chemical composition and input physics.
We are thus able to compare very easily the `shape'
of the growth rate evolution for models that are rather different
from each other. For overtone modes higher than the 1O,
the trend is almost the same for all masses,
the major effect being a systematic (but small) decrease in growth
rates towards the high masses. This is highlighted
in the right panels of Fig.~\ref{fig:GR_dep_M},
where the value $GR_{\rm max,n}$ of the growth
rate at the cut-off is displayed as a function of mass.
For the 1O mode, growth rates are substantially
larger towards higher masses, especially in
the neighbourhood of the acoustic cut-off.
On the other hand, at small values of $\tilde{\nu}$,
more massive models have smaller growth rates,
possibly negative.

Fig.~\ref{fig:GR_dep_M} also shows the fundamental mode
growth rates for which $\tilde{\nu}$ is less effective
in providing a framework for comparison.
The reason, as already discussed, is that
overtone modes are largely determined by surface properties,
encoded by $\nuac$, while the fundamental mode
is more sensitive to the envelope structure.
The fundamental mode experiences temporary stabilisation
only for a limited interval of masses,
$1.2\lesssim M/\,\Msun\lesssim 3.5$. This is especially
clear from the top-right panel of Fig.~\ref{fig:GR_dep_M},
showing the value of $GR_{\rm min,0}$ (the fundamental
mode growth rate at the local minimum)
as a function of mass. Outside that range of masses,
we find two possible trends: either the growth rates experience a positive minimum
thus not becoming stable, or show an essentially monotonic
increases with no minimum at all. The latter occurs for
stars more massive than $\sim4.5\,\Msun$. Note that
the most massive models also show an earlier rise of
the growth rate (at lower values of $\tilde{\nu}$ or $L$).
A consequence of these trends is that, in the most
massive models, it is possible for the fundamental mode
to be excited and dominant over the whole luminosity interval of the grid.
In that case, the regular pattern of shifts of the dominant mode towards
lower radial orders, discussed in the previous sections, does not occur.

\subsubsection{Varying metallicity}
The effect of varying metallicity on the behaviour
of growth rates is displayed in Fig.~\ref{fig:GR_dep_Z},
and is essentially negligible for the 3O and 4O modes.
The only relevant effect for the 2O mode occurs
when it has already exceeded the acoustic cut-off,
and is thus not relevant. On the other hand, we find
the 1O mode growth rate at the cut-off to be
significantly larger at low metallicity. This
effect is even more pronounced at higher luminosities.
This trend is found also for the fundamental mode,
that, in the regime of temporary stabilisation,
shows larger growth rates for lower values of metallicity.

\subsubsection{Varying helium and hydrogen abundances}
Fig.~\ref{fig:GR_dep_XY} compares the evolution of
growth rates for series of models with different
values of the helium mass fraction, $Y$.
Again, we find mild effects on the
values of the growth rates of the 2O, 3O, and 4O modes,
while the 1O mode shows larger values of $GR_{\rm max,1}$
for higher helium contents. Note, however, that
this occurs as the more He-rich sequences reach
the acoustic cut-off later (the peak of growth
rates appears at larger $\tilde{\nu}$ for all overtones).
This is a consequence of a higher value of the
mean molecular weight term in Eq.~\ref{eq:nuac_isothermal},
neglected in the definition of $\tilde{\nu}$ (Eq.~\ref{eq:tildenu}).
As for the fundamental mode, a higher abundance of helium
adds a significant contribution to the driving,
especially in the regime of temporary stabilisation.

\subsubsection{Varying $\co$}
\begin{figure}
    \includegraphics[width=\columnwidth]{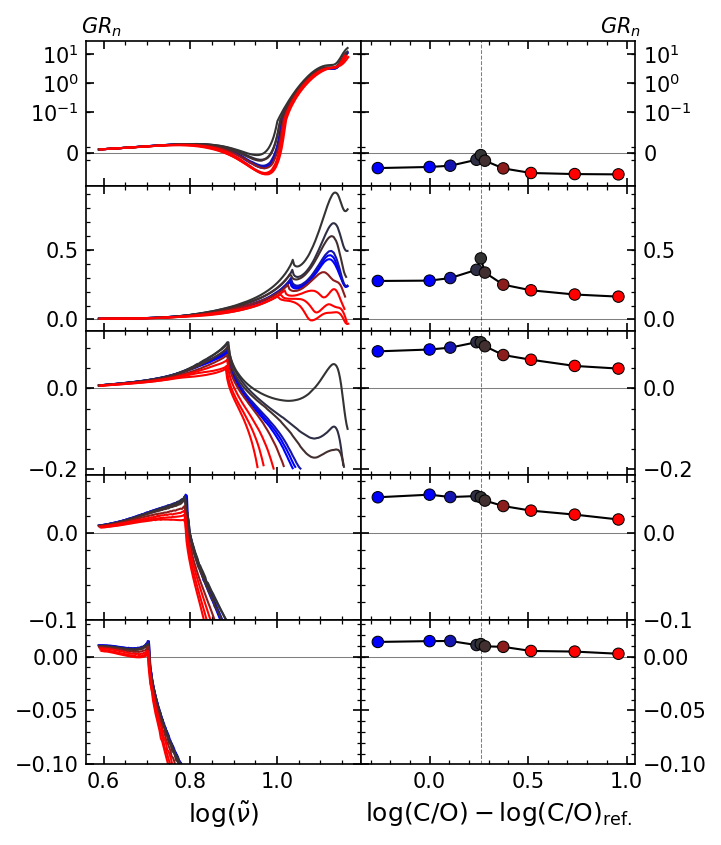}
    \caption{Same as Fig.~\ref{fig:GR_dep_M}, but showing series
    of models with different $\co$.
    Other global parameters are those summarised in Table~\ref{tab:SelectedSequence}.
    Vertical dashed lines in the right panels
    indicate $\co=1$.}
    \label{fig:GR_dep_fCO}
\end{figure}

The impact of an enhanced carbon abundance on
growth rates is shown in Fig.~\ref{fig:GR_dep_fCO}.
In the regime where overtone modes higher than the 2O
are excited, the surface temperature of the models
is too large for the formation of molecules
to have any appreciable effect on pulsation.
Thus, the decrease of $GR_{\rm max,n}$
towards higher $\co$ simply reflects the
effect of an increased metallicity.
For the 2O and 1O modes, stronger effects begin to be visible,
with growth rates being enhanced in the vicinity
of $\co\simeq1$. The same is true for the fundamental
mode, in the neighbourhood of the point of minimum
growth rate. Note that the growth rate peak
of the 1O mode shows the same displacement discussed
in the case of varying He abundance, though in
this case the cause is mostly the variation of the $\Gamma_1$
term in Eq.~\ref{eq:nuac_isothermal} rather than
the mean molecular weight term.

\subsubsection{Varying $\aml$}
\begin{figure}
    \includegraphics[width=\columnwidth]{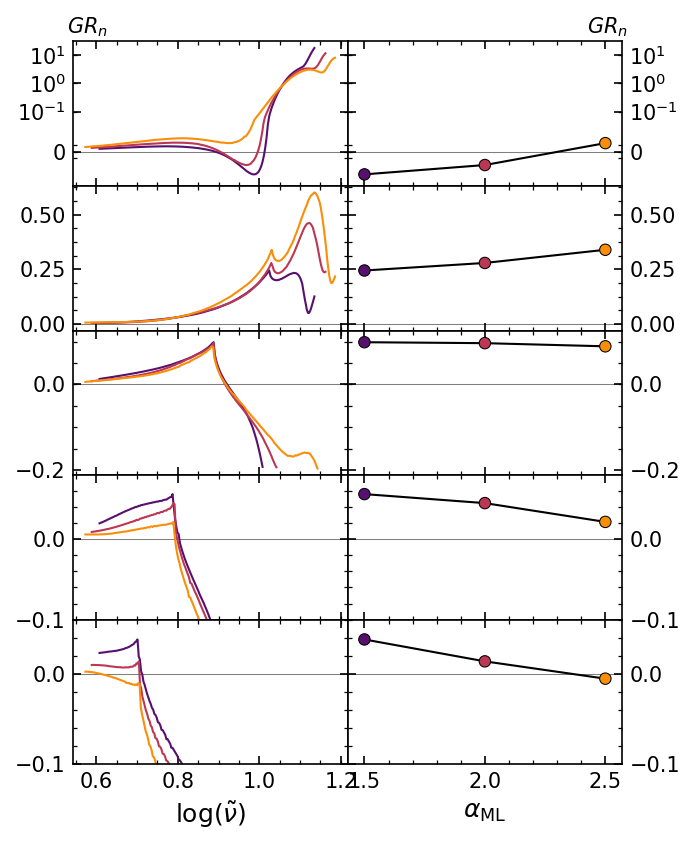}
    \caption{Same as Fig.~\ref{fig:GR_dep_M}, but showing series
    of models computed with a different value of the mixing length parameter $\aml$.
    Other global parameters are those summarised in Table~\ref{tab:SelectedSequence}.}
    \label{fig:GR_dep_aML}
\end{figure}

The growth rates of low- and high-order modes show an opposite
dependence upon the mixing length parameter (Fig.~\ref{fig:GR_dep_aML}).
The 3O and 4O modes suffer a significant reduction in growth rates
when $\aml$ is increased, while the opposite is
true for the 1O and fundamental mode. The 2O mode
growth rate is remarkably insensitive to the
value of $\aml$, at least for the combination
of mass and chemical composition considered here.
Note that the effect on the fundamental mode
growth rates resulting from increasing $\aml$
is somewhat similar to that produced by an
increase in helium content.

\subsection{Analytic prescription for pulsational stability}
\label{sec:StabilityFit}

The regularity shown by overtone modes growth
rates allows us to construct a conveniently simple model
to describe pulsational stability as a function of stellar parameters.
To begin with, we recall that overtone modes become
stable when the star evolves to the point where
their frequency exceeds the acoustic cut-off value,
determined by surface properties.
Usually, this results in the exclusion of all overtone
modes with a radial order larger than some value $n_d$.
Among the remaining modes, the one with $n=n_d$
has the shortest period, and very likely the
largest growth rate, so we can expect it to be dominant.
Modes with $n<n_d$ are likely excited, the mode
with $n=n_d-1$, having the second-largest growth rate,
followed by the mode with radial order $n_d-2$, and so on.

Since the fundamental mode properties depend
intrinsically on the interior structure, an equivalent
modelling is not as simple. However, models show
that it usually becomes dominant when the 1O mode
has already exceeded the acoustic cut-off,
at least for $M\lesssim3\,\Msun$, so that
it fits the scheme discussed above in most cases of interest.

The main difficulty in applying the above algorithm
resides in the computation of $\nuac$.
If the approximation of isothermal atmosphere
is valid, it can be computed with Eq.~\ref{eq:nuac_isothermal},
using the values of $M$, $R$, $T$ at the surface.
Note, however, that the mechanical boundary
condition in the present pulsation model is
not defined at the photosphere, but rather at
the outermost point of the models, that can
be significantly far from $R_{\rm phot}$
(see, \eg, the bottom panels of Fig.~\ref{fig:SelectedFundMode_W_G1}).
As an additional complication, a knowledge
of the surface values of $\mu$ and $\Gamma_1$
is in principle required to employ Eq.~\ref{eq:nuac_isothermal}.
In general, one can assume that $\mu\sim\Gamma_1\sim1$,
but the approximation is not necessarily good
for evolved TP-AGB stars that experience
significant changes in composition and
molecule formation in the atmosphere.

In order to provide a more easily applicable
prescription, we formulate the problem in
terms of the critical luminosity $L_{{\rm ac},n}$
at which an overtone mode of given radial order $n$
reaches the acoustic cut-off frequency, for
given mass and chemical composition.
This luminosity can be interpreted as
the transition luminosity between dominant
pulsation in the modes $n$ and $n-1$.
The critical luminosity can be expressed as:

\begin{align}\label{eq:logLaclogM}
    & \log \left( \frac{L_{{\rm ac},n}}{\Lsun} \right) = a_{\rm 0,n} + a_{\rm Z,n} \log(Z) + a_{\rm X,n} X + \\
    & +a_{\rm \co,n} \log\left[\frac{\co}{(\co)_{\odot}}\right] + \left[ b_{\rm 0,n} + b_{\rm X,n} X \right] \log \left( \frac{M}{\Msun} \right) \,. \nonumber
\end{align}

The coefficients for different radial orders $n$ are
shown in Table~\ref{tab:logLaclogM_coeff_Orich} for the O-rich case
and in Table~\ref{tab:logLaclogM_coeff_Crich} for the C-rich case.
For O-rich models, $L_{\rm ac}$ does not depend on $\co$ ($a_{\rm \co,n}=0$).
For C-rich models, the effect of $\co$ is appreciable only at low
enough values of $\Teff$ to allow for significant molecule formation
in the atmosphere. The 3O and 4O modes generally reach the acoustic
cut-off at relatively small radii and warm surface temperature,
so that $L_{{\rm ac},n}$ for $n=3,4$ is essentially insensitive to $\co$
even in C-rich models. We thus provide coefficients of Eq.~\ref{eq:logLaclogM}
for C-rich composition only for $n=1,2$.

\begin{table}
\normalsize
\centering
\caption{
    Coefficients of Eq.~\ref{eq:logLaclogM} for the four lowest-order overtone modes,
    for the O-rich case (in which case $a_{\co}=0$).
    }
\label{tab:logLaclogM_coeff_Orich}
\begin{tabular}{c||c|c|c|c}
 $n$ & 1 & 2 & 3 & 4 \\
\hline
$a_0$       & +4.0107 & +3.5724 & +3.2324 & +2.9626 \\
$a_Z$       & -0.1466 & -0.0874 & -0.1036 & -0.1013 \\
$a_X$       & -0.9965 & -0.8453 & -0.8789 & -0.9045 \\
$b_0$       & +1.5521 & +1.9692 & +1.7961 & +1.8182 \\
$b_X$       & +0.5154 & -0.0715 & +0.1301 & +0.1424 \\
\end{tabular}

\caption{
    Coefficients of Eq.~\ref{eq:logLaclogM} for the first and second overtone modes, for the C-rich case.
    }
\label{tab:logLaclogM_coeff_Crich}
\begin{tabular}{c||c|c}
 $n$ & 1 & 2 \\
\hline
$a_0$       & +3.8474 & +3.3819 \\
$a_Z$       & -0.2079 & -0.1486 \\
$a_X$       & -1.0222 & -0.7945 \\
$a_{\co}$   & -0.1613 & -0.0514 \\
$b_0$       & +1.5895 & +1.9311 \\
$b_X$       & +0.4440 & -0.0471 \\
\end{tabular}
\end{table}

One can use Eq.~\ref{eq:logLaclogM} to compute the luminosity
at which a given overtone mode would reach the acoustic cut-off
for a given stellar model. By computing it for $n=1$ to $4$,
and comparing the results with the actual luminosity of the model,
one is able to tell that all modes for which $L>L_{{\rm ac},n}$ are
stable and not observable, the highest overtone mode (of radial order, say, $n=n_d$)
for which $L<L_{{\rm ac},n}$ is the dominant, and the remaining modes are
likely excited and have ordered growth rates such that $GR_{n_d}>GR_{n_d-1}>GR_{n_d-2}>...$, and so on.
If $L>L_{{\rm ac},1}$, all overtone modes are stable and the fundamental mode is dominant.
We recall that this approach is valid only for $M\lesssim3\,\Msun$,
above which value one expects the fundamental mode to be
dominant at all luminosities in the interval explored here.

\section{Pulsation Periods}
\label{sec:Periods}

\subsection{Dependence of periods upon stellar parameters}
\label{sec:DependencePeriodsStellarParameters}
\begin{figure}
\centering
    \includegraphics[width=\columnwidth]{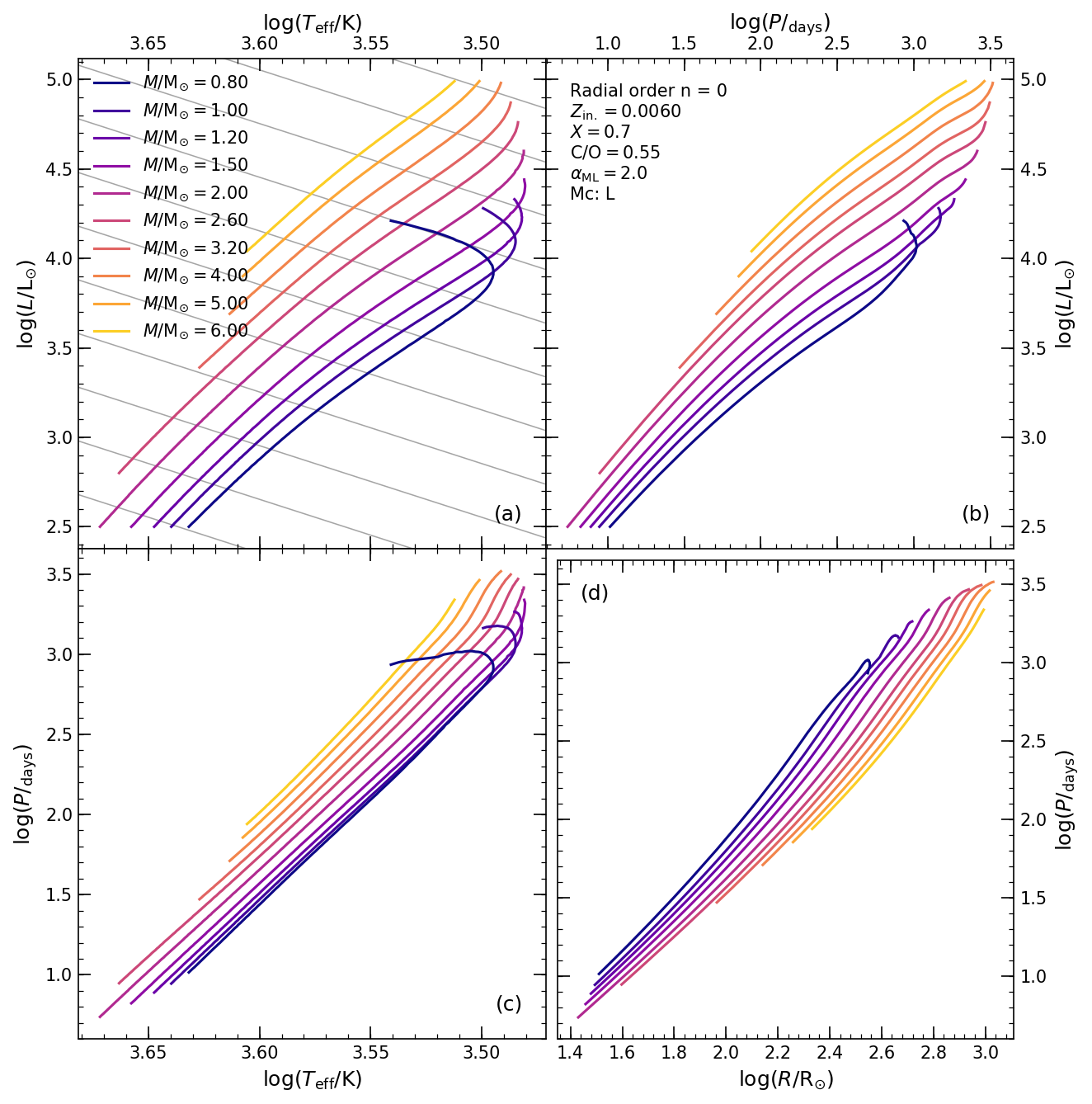}
    \caption{Series of models with $Z=0.006$, $X=0.7$, $\co=(\co)_{\odot}\simeq0.55$, $\aml=2$
    and core mass computed with the `lower limit' CMLR. Models are shown for masses
    $0.8$, $1.0$, $1.2$, $1.5$, $2.0$, $2.6$, $3.2$ and $4.0\,\Msun$ (colour coded).
    Panel (a) shows the models in the HR diagram, while the fundamental mode period $P_0$
    is shown against luminosity (panel (b)), effective temperature (panel (c))
    and photospheric radius (panel (d)). Lines of equal radius are shown in the background of panel (a).
    }\label{fig:P0_var_M}
\end{figure}

In the present section we discuss the dependence
of pulsation periods upon global stellar parameters,
focusing on the fundamental mode and on its dependence
upon the stellar radius. The latter aspect
is particularly interesting since,
to a good degree of approximation,
pulsation periods are determined
by mass and radius via the
period-mean density relation.
This is commonly expressed in
terms of a period-mass-radius (PMR) relation:
\begin{equation}\label{eq:logPMR}
    \log(P/{\rm days}) = a_0
                       + a_M\,\log\left(\frac{M}{\Msun}\right)
                       + a_R\,\log\left(\frac{R}{\Rsun}\right) \,.
\end{equation}
Eq.~\ref{eq:logPMR} provides a convenient
mean to predict periods for stellar models
given their mass and radius, or to estimate
stellar masses based on observed periods
and an independent measurement of the stellar radius.

A significant advantage of this approximation
is that it is essentially independent of
other global stellar properties.
In fact, changes in the chemical composition
of the envelope have the main effect of
changing the effective temperature by
modifying the opacity sources, and thus
their effect is already incorporated
in the stellar radius. Similarly,
differences in core mass result
in different luminosities, and again
the effect is already included in the stellar radius.

Having computed pulsation models for a wide
range of stellar parameters, we are able
to verify the degree to which this approximation is valid.
We find it to be very well verified for overtone
mode periods, with deviations safely within $\sim1$ per cent.
The approximation is also generally valid
for the fundamental mode, though with larger
deviations, up to $\sim20$ per cent. To show this,
we examine the differences in periods,
at fixed mass and radius, resulting from changing
one other global parameter at a time.
In particular, we study the effects of
changing the metallicity, the abundance of helium,
and the $\co$ ratio. Additionally we consider
the effects produced by different choices
of the mixing length parameter $\aml$ in the calculation of models.

Fig.~\ref{fig:P0_var_M} shows the dependence of the fundamental
mode period on radius, luminosity and effective
temperature at constant mass, for series
of models of several masses. The Hertzsprung-Russel
diagram (HRD) for those models is also shown.
The interpretation of the diagram is quite
straightforward using Eq.~\ref{eq:logPMR}
and recalling that $L\propto R^2\Teff^4$. At a given luminosity,
the more massive models have higher effective
temperature (\ie, warmer Hayashi lines, panel (a)),
therefore they have smaller radii and shorter periods (panel (b)).
Conversely, at a given effective temperature,
more massive models are brighter, have
larger radii and longer periods (panel (c)).
Panel (d) shows how models with different masses
follow different period-radius relations, in accordance with Eq.~\ref{eq:logPMR}.

\subsubsection{Varying metallicity}
\begin{figure}
\centering
    \includegraphics[width=\columnwidth]{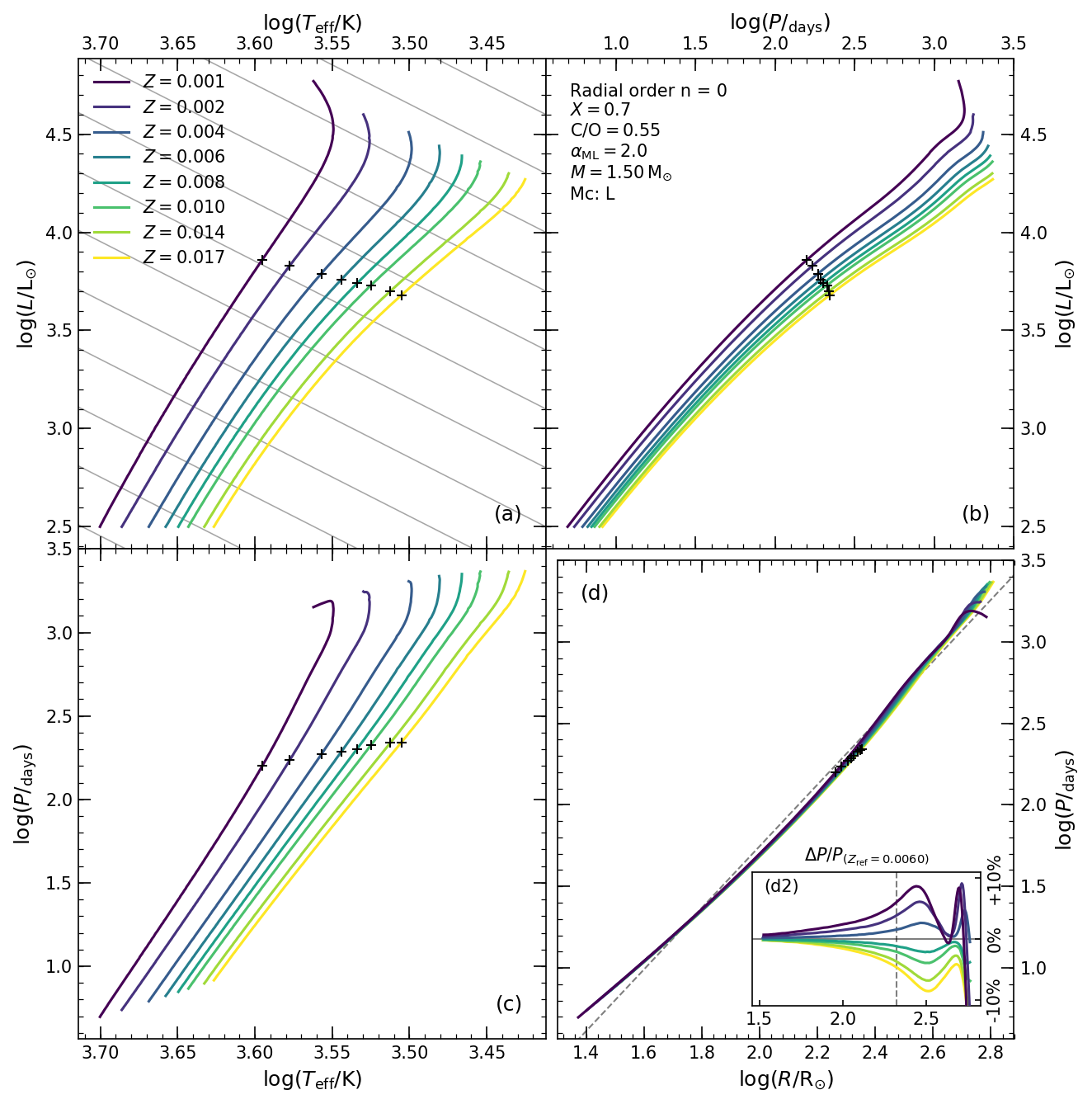}
    \caption{Same as Fig.~\ref{fig:P0_var_M}, but only for $M=1.5\,\Msun$,
    and displaying series of models with different values of metallicity,
    (colour coded). The period-radius relations are
    almost the same, regardless of $Z$.
    Inset panel (d2) shows the relative difference in period
    (with respect to the case $Z=0.006$ taken as a reference) as a function of $\log(R/\Rsun)$.
    Plus symbols mark, along each series of models,
    the approximate point where the fundamental mode
    is expected to become excited to a clearly detectable level.
    The vertical dashed line in panel (d2) has the same meaning.
    }\label{fig:P0_var_Z}
\end{figure}

\begin{figure}
\centering
    \includegraphics[width=\columnwidth]{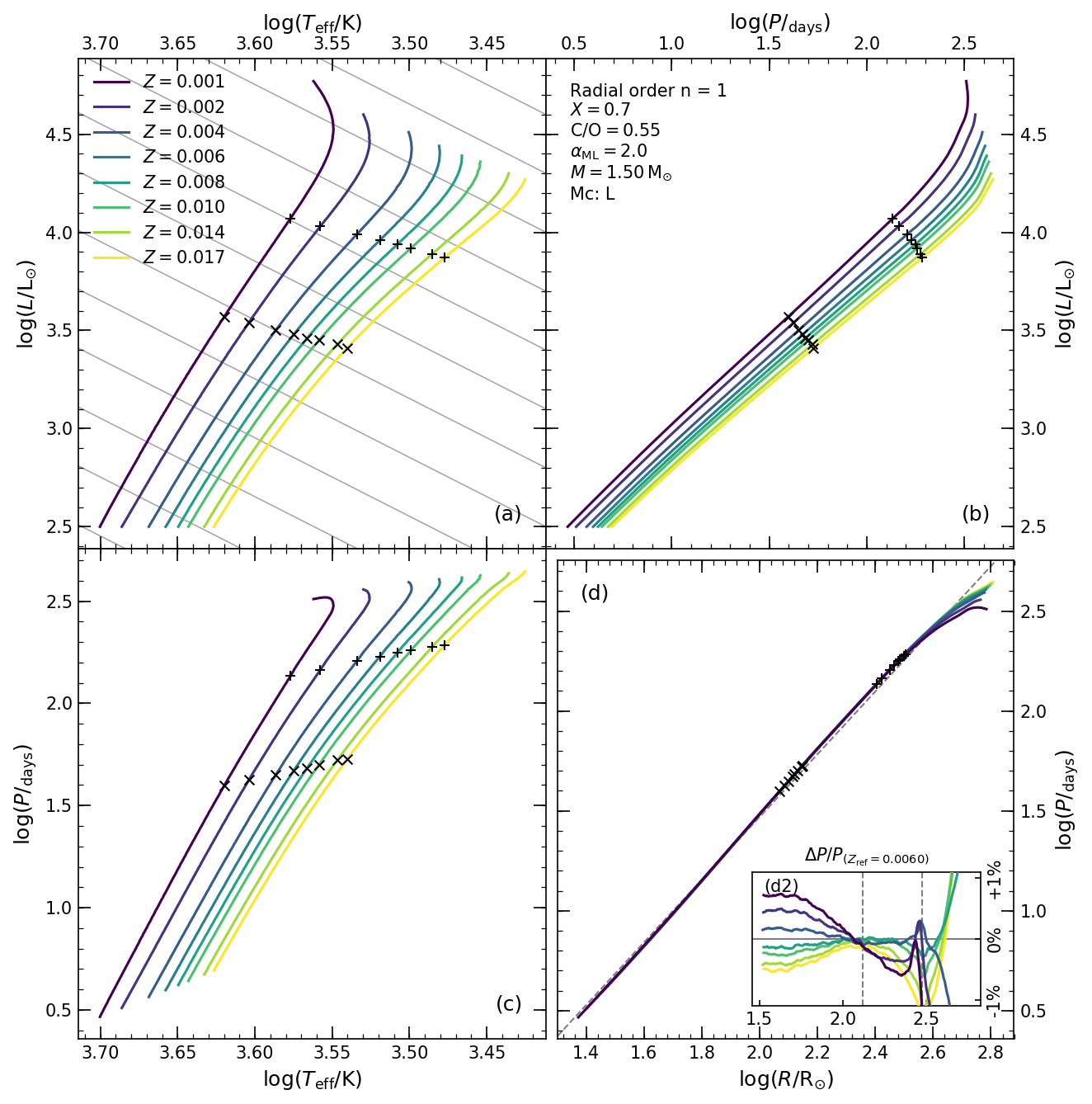}
    \caption{Same as Fig.~\ref{fig:P0_var_Z}, but for the 1O mode.
    }\label{fig:P1_var_Z}
\end{figure}

If mass is kept fixed, from Eq.~\ref{eq:logPMR}
we expect periods to follow the same relation
with radius regardless of other stellar properties,
such as chemical composition.
This is displayed in Fig.~\ref{fig:P0_var_Z},
showing the effect of varying the
envelope metallicity $Z$ when the other
parameters are kept fixed
(except for the mass fraction of helium,
that varies accordingly with $Z$ since $X$ is fixed).
In the HRD, the more metal-poor series of models
have higher $\Teff$ at a given luminosity,
implying a shorter period. As expected,
the effect is largely removed when periods
are shown against the stellar radius (panel (d)),
in which case all series of models lie very close to
each other. However, a close inspection
reveals that the difference in period
between the most metal-rich sequence ($Z=0.017$)
and the most metal-poor one ($Z=0.001$)
can become as large as $20$ per cent even though
models have the same mass and radius.
This is shown in the inset (d2) of panel (d),
where the relative period difference
(with respect to the series of models
with $Z=0.006$, taken as reference)
is displayed as a function of radius.
In the case of the $1.5\,\Msun$ models
in Fig.~\ref{fig:P0_var_Z}, the largest differences
occur at $R\sim250\,\Rsun$ and
$300\lesssim P/{\rm days}\lesssim500$,
the range of periods observed 
in Miras and fundamental mode SRVs.
Such differences in period are due to
the fact that the coefficients $a_M$, $a_R$
in Eq.~\ref{eq:logPMR} are not actually
independent of mass and radius.
This is discussed in more detail in
Sect.~\ref{sec:PeriodRadiusRelation}.
In contrast, the relation between period
and radius for overtone modes is much
more regular, resulting in minute deviations.
The largest differences of period at equal
radius occur for the 1O mode, for which
they are generally smaller than $\sim1$ per cent
(as shown in Fig.~\ref{fig:P1_var_Z}
for varying metallicity).

\subsubsection{Varying helium and hydrogen abundances}
\begin{figure}
\centering
    \includegraphics[width=\columnwidth]{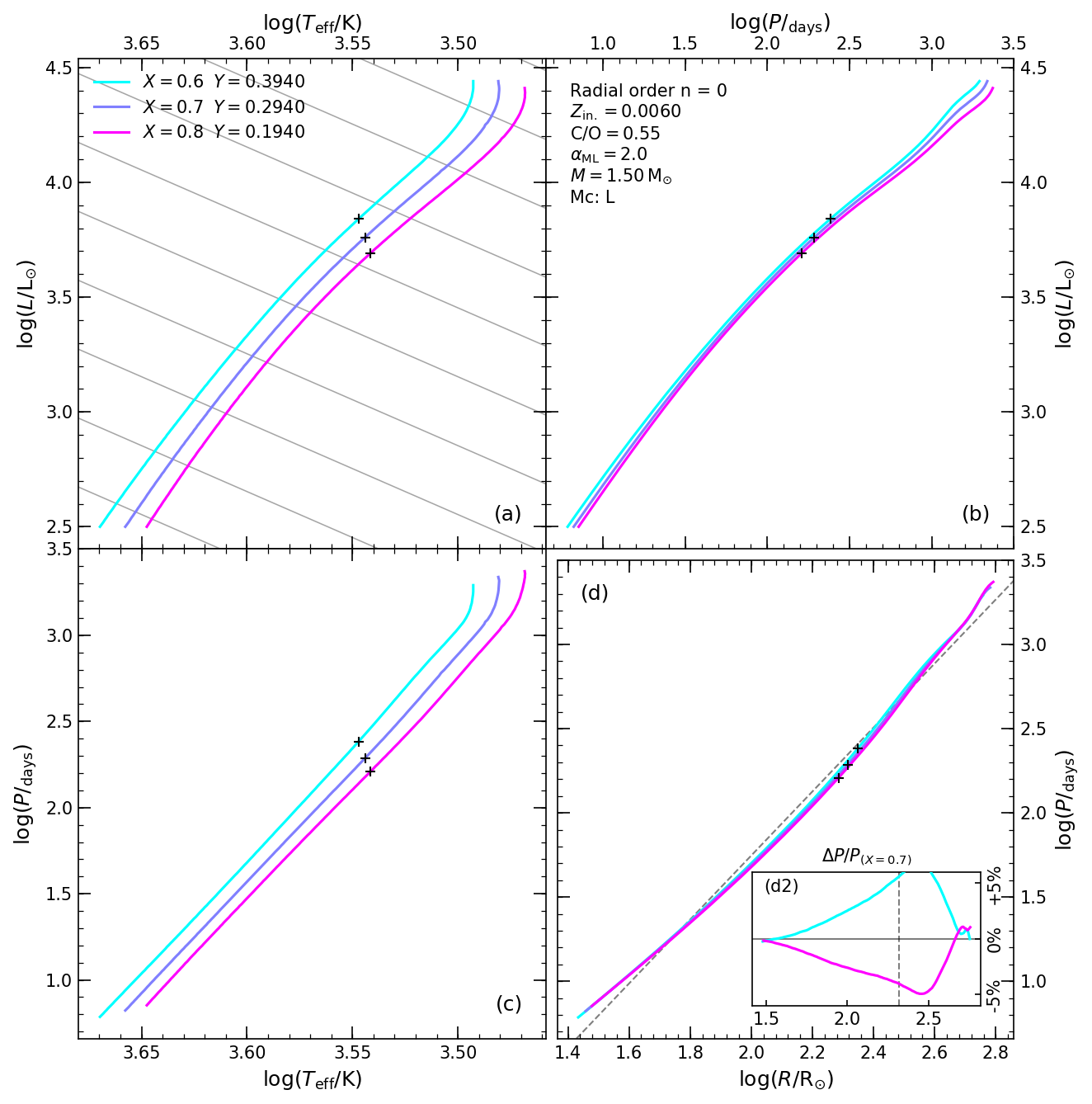}
    \caption{Same as Fig.~\ref{fig:P0_var_Z}, but with series of models differing
    in the mass fraction of hydrogen (colour coded).
    }\label{fig:P0_var_X}
\end{figure}

Similar considerations, regarding the fundamental mode period,
can be made when changing the abundances of hydrogen and helium.
Three cases of this situation are considered in the grid of
models, obtained with three different values of the hydrogen
mass fraction, $X=0.6, 0.7,$ and $0.8$. Each case correspond
to a different value of He mass fraction as well, determined
by $Y=1-X-Z$, so that the abudance of helium actually depends
upon metallicity in the models. Fig.~\ref{fig:P0_var_X}
shows the case for $Z=0.006$ in which $Y=0.194, 0.294,$ and $0.394$.
The effect is qualitatively similar to, but less pronounced than,
that produced by changing $Z$,
despite of the rather substantial difference $\Delta Y=0.2$
under consideration.

\begin{figure}
\centering
    \includegraphics[width=\columnwidth]{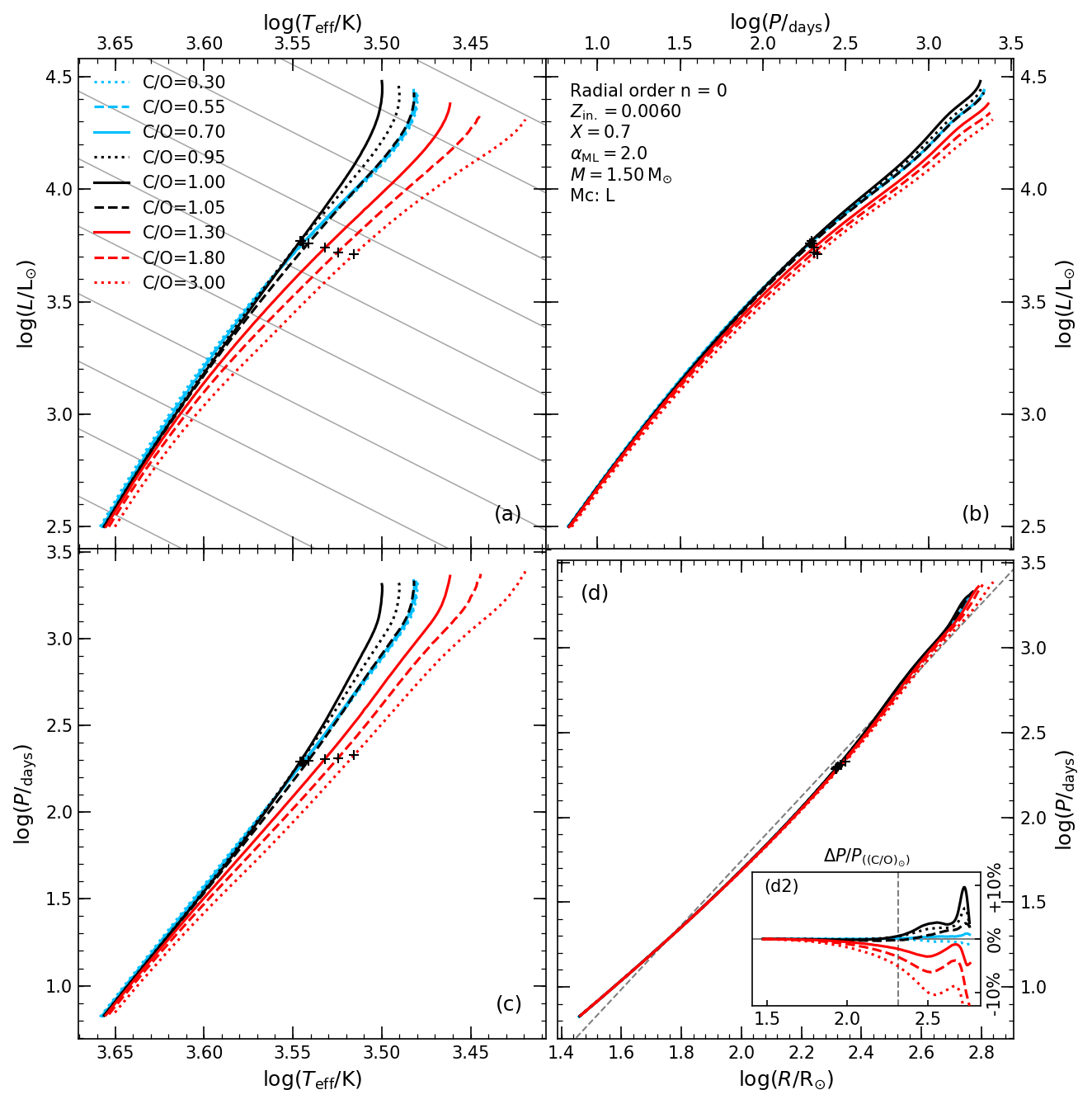}
    \caption{Same as Fig.~\ref{fig:P0_var_Z},
    but with series of models differing in the $\co$ (colour coded).
    }\label{fig:P0_var_CO}
\end{figure}

\begin{figure}
\centering
    \includegraphics[width=\columnwidth]{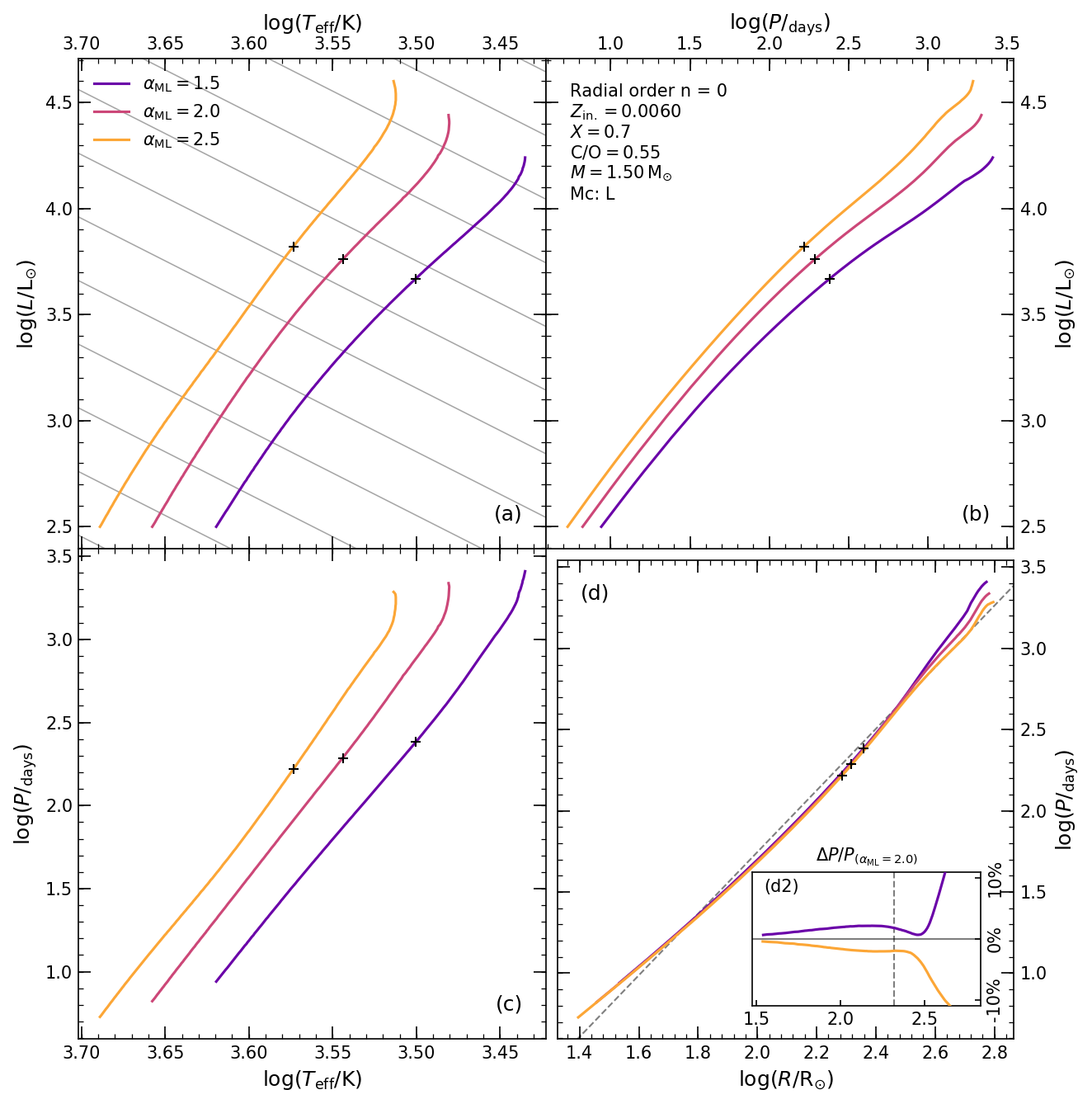}
    \caption{Same as Fig.~\ref{fig:P0_var_Z}, but with series of models differing
    in the value of the mixing length parameter $\aml$ used in the computation (colour coded).
    }\label{fig:P0_var_aML}
\end{figure}

\subsubsection{Varying $\co$ ratio}
The case of varying $\co$ (Fig.~\ref{fig:P0_var_CO}) is particularly
interesting. During the TP-AGB
phase, repeated third dredge-up events
bring to the surface the products of
nuclear burning, enriching the envelope
in carbon, and increasing the $\co$
to values possibly larger than 1, and thus
resulting in the formation of carbon stars.
When this happens, the sources of molecular
opacities in the atmosphere are drastically
altered. In fact, owing to the high binding
energy of the CO molecule, essentially all
carbon atoms in the atmospheres of O-rich
stars are locked to form CO, and the low
temperature opacity is dominated by
oxygen molecules such as TiO, VO, H$_2$O.
Conversely, the spectra of carbon stars
have strong absorption features of C-bearing
molecules (C$_2$, CN, SiC) while
all the oxygen atoms are locked to form CO
\citep[\eg,][]{Marigo_Aringer_2009}.

The net result is that, at a given luminosity,
C-stars have systematically lower
effective temperatures than O-rich stars,
and thus larger photospheric radii.
This can be seen in the HRD (panel (a)) of Fig.~\ref{fig:P0_var_CO},
where series of models of different $\co$
are compared. The difference is most evident
at high luminosity since the $\Teff$ is lower
and it is easier to form molecules.
It is interesting to observe that the effect
of varying $\co$ is essentially non-existent
as long as $\co\lesssim0.95$, while the largest
temperatures are attained in models with $\co\simeq1$
(in which all C atoms as well as the O ones
are locked into CO), causing shorter periods.
At a given luminosity, models with higher $\co$
are colder and have longer periods.
Conversely, at a given $\Teff$ they
are less bright and have shorter periods.

At given radius, the relative differences in period
are generally small (less than $\sim0.5$ per cent)
when $\co\lesssim1$, while the most C-rich
models have periods up to $10$ per cent smaller than
those of O-rich stars.

\subsubsection{Varying $\aml$}
Finally, we consider the cases of series
of models computed with different values
of the mixing length parameter, Fig.~\ref{fig:P0_var_aML}.
Models computed with $\aml=1.5$ are 20\% 
 cooler than those in which
$\aml=2.5$ was used, but the effect is
largely incorporated in differences
in radius, so that at the same radius
the fundamental mode periods do not differ
by more than $\sim5$ per cent (having almost doubled $\aml$).
Note that the large deviations at $\log(R/\Rsun)>2.5$
(Fig.~\ref{fig:P0_var_aML}, panel (d2)) occur
as series of models obtained with a different
value of $\aml$ leave the Hayashi line
at different radii.

\subsection{The period-radius relation of the fundamental mode}
\label{sec:PeriodRadiusRelation}

In the previous section we have shown that
for overtone modes, theoretical period-radius relations
at fixed mass generally fulfil the expectation of
being independent of chemical composition and
other input physics, while the same is not true
for the fundamental mode period. Again, this
is a consequence of the relatively low sensitivity
of the properties of overtone modes to
the interior structure of the models.
In contrast, the fundamental mode is substantially
affected by the conditions in the envelope's interior
\citep[see][in particular their fig.~2]{Fox_Wood_1982}.
The details of such dependence will be addressed in
a forthcoming study. Here, we only point out
that a PMR relation in the form Eq.~\ref{eq:logPMR} is
not appropriate to describe the period of the fundamental mode.

\begin{figure}
    \includegraphics[width=\columnwidth]{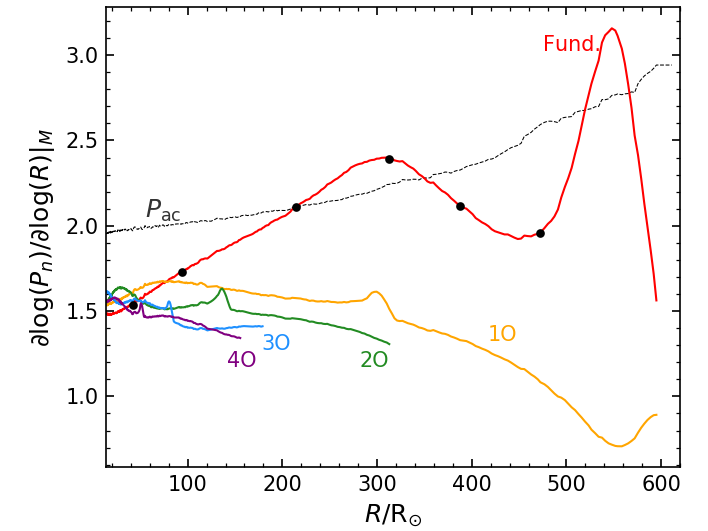}
    \caption{Derivative $\partial\log(P_n/{\rm days})/\partial\log(R/\Rsun)$
    as a function of radius for modes in the selected sequence
    (see Fig.~\ref{fig:SelectedSequence} and Table~\ref{tab:SelectedSequence}).
    Bumps along lines of overtone modes correspond to the acoustic cut-off,
    and the dashed line is the derivative of the acoustic cut-off period, $P_{\rm ac}=1/\nuac$.
    }
    \label{fig:Selected_dPdR}
\end{figure}

In fact, the dependence
of the fundamental mode period upon stellar
radius in the models can be described as a
power-law only to a first approximation
(in other words, $\log(P_0)$ is not exactly
linear in $\log(R/\Rsun)$). Similar considerations
can be made for the dependence upon mass.
If a functional form such as Eq.~\ref{eq:logPMR}
is used to describe fundamental mode periods,
the coefficients $a_M$ and $a_R$ themselves
should depend upon mass and radius.
In the case of radius dependence, this is
evident from Fig.~\ref{fig:Selected_dPdR}, where the local
derivative $\overline{a}_{R,n}=\partial\log(P_n)/\partial\log(R)|_M$
of period with respect to radius, at constant mass,
is shown for each mode as a function of stellar radius.
For overtone modes, $\overline{a}_{R,n}\simeq1.5$
prior to the cut-off, consistently with the
period-mean density relation ($P\propto M^{-1/2}R^{3/2}$).
For the fundamental mode, $\overline{a}_{R,0}=1.5$
at small radii, but then it increases linearly
with radius. The trend is not monotonic beyond $R\simeq300\,\Rsun$.%but rather
%alternates with phases in which $\overline{a}_{R,0}$
%decreases, also linearly, with radius.
Note that $\overline{a}_{R,0}$ increases substantially
at large radii, doubling its original value.

This dependence of the fundamental
period upon radius has two consequences.
Firstly, it is the very reason for
it never reaching the acoustic cut-off
(cf. Fig.~\ref{fig:SelectedSequence}, top panel;
see also the discussion in Sect.~\ref{sec:StabilityFundamental}).
Secondly, it produces the deviations described in
Sect.~\ref{sec:DependencePeriodsStellarParameters}.
The latter fact becomes clear by looking at inset (d2)
of Fig.~\ref{fig:P0_var_Z} (especially in comparison
with Fig.~\ref{fig:P1_var_Z}): the largest deviations
are due to the fact that the same trend of $\overline{a}_{R,0}$
occurs for all metallicities, but at different radii.

\subsection{Fitting relations for periods}
\label{sec:PeriodsFit}

\begin{figure}
    \includegraphics[width=\columnwidth]{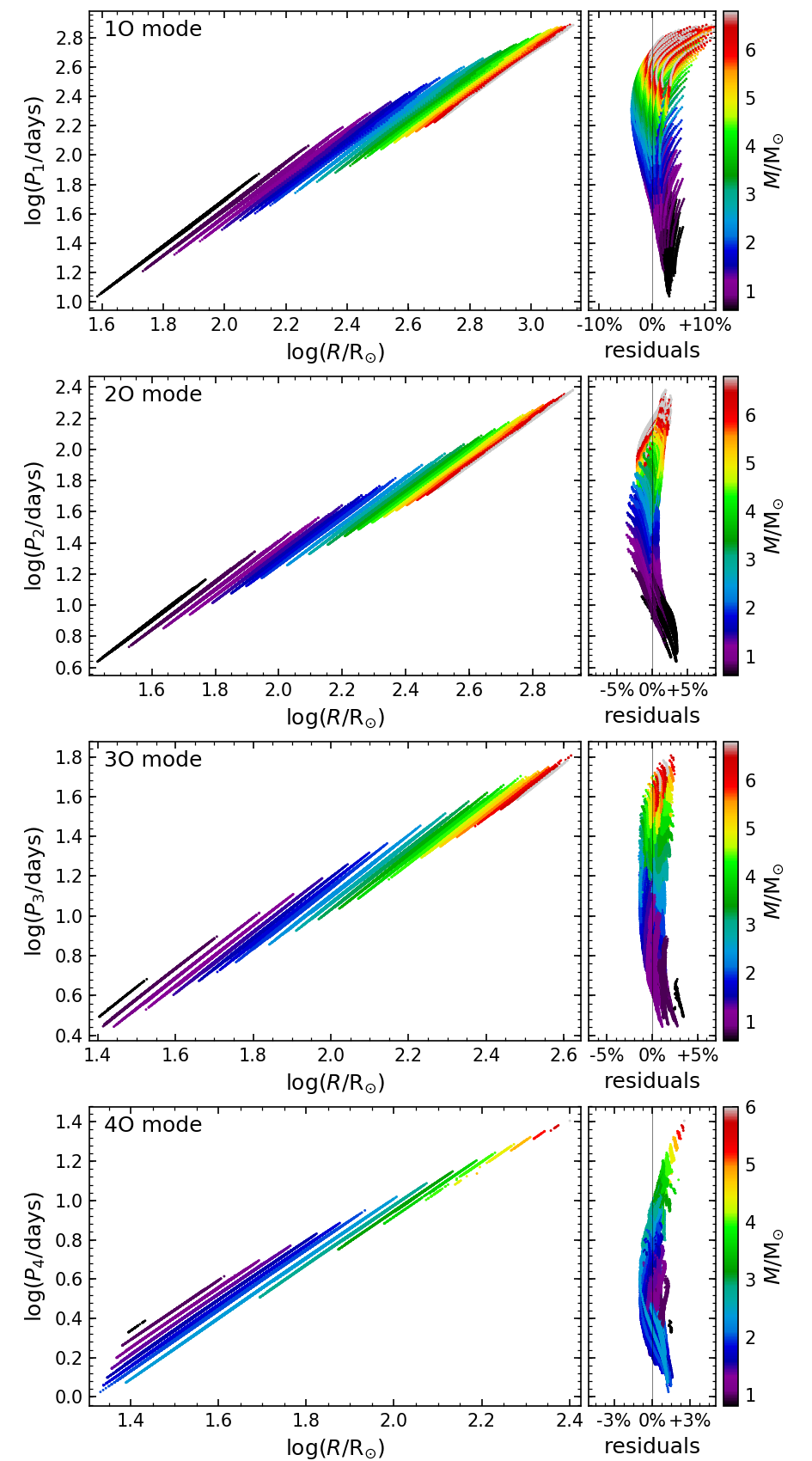}
    \caption{Period-mass-radius relation for overtone modes (left),
    and residuals of the fitting relations (right). For each overtone $n$,
    the plots are limited to models for which $GR_n>0$ and the
    mode is dominant based on Eq.~\ref{eq:logLaclogM}.}
    \label{fig:FitOvertonePeriods}
\end{figure}

We provide here best-fit coefficients
for Eq.~\ref{eq:logPMR} to compute pulsation
periods of overtone modes as a function of
mass and radius (Table~\ref{tab:PMRcoefficientsOvertones}).
Confirming the discussion in the previous
sections, the PMR in the form Eq.~\ref{eq:logPMR}
allows for a good prediction of overtone
mode periods, reproducing values from the
models to within $\sim10$ per cent for the 1O mode,
and to within $\sim5$ per cent or less for higher overtones
(see Fig.~\ref{fig:FitOvertonePeriods}).
For each mode, we limited the best-fit derivation to
models in which that mode is dominant
according to the prescription provided
in Sect.~\ref{sec:StabilityFit}, Eq.~\ref{eq:logLaclogM}.
Outside such limits, predictions from
the best-fit relations are not necessarily as good.

\begin{figure}
    \includegraphics[width=\columnwidth]{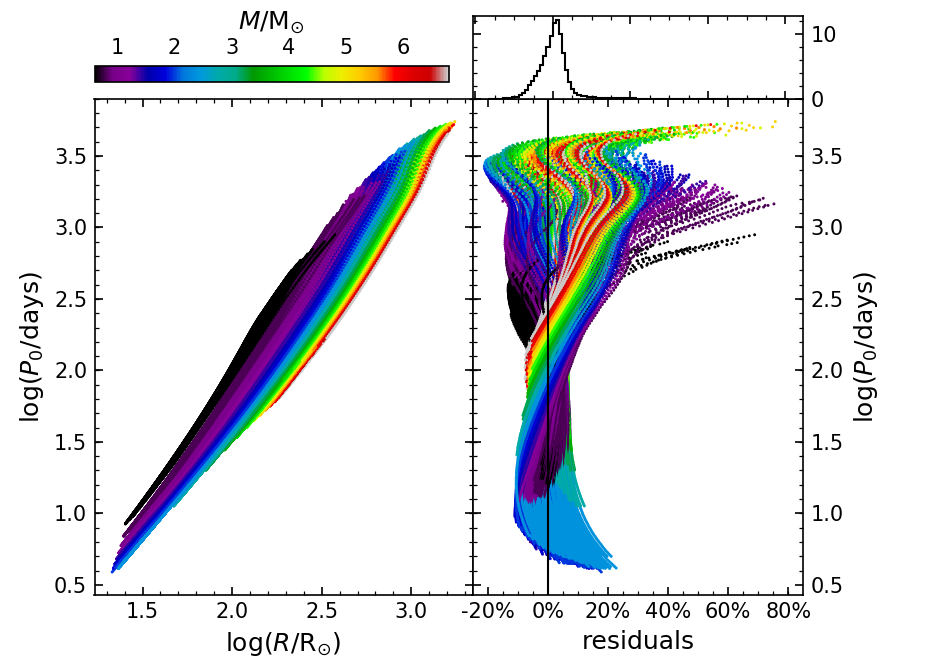}
    \caption{Period-mass-radius relation for the fundamental mode (left),
    and residuals of the fitting relation (right). Only models
    which $\Teff$ is decreasing with luminosity are shown.}
    \label{fig:FitFundamentalPeriod}
\end{figure}

Unfortunately, a similarly effective and compact
description of the fundamental mode is not possible,
for the reasons explained in Sect.~\ref{sec:PeriodRadiusRelation}.
Also, it is not trivial
to limit the derivation of the best-fit coefficients
to a simply defined range of stellar parameters.
Here, we provide a reasonable compromise with
an expression not too complex but still
able to capture the general trend of the models.
Since a linear dependence upon $\log(M/\Msun)$ and $\log(R/\Rsun)$
is not valid on the whole range of parameters,
we opted for a bi-cubic form instead.
Additionally, we included the dependence upon
metallicity, helium mass fraction, and $\co$:
\begin{align}
    & \log(P) = a_0 + a_M\tilde{m} + a_R\tilde{r}
    + b_{M}\tilde{m}^2 + b_{MR}\tilde{m}\tilde{r} + b_{R}\tilde{r}^2 \nonumber \\
    & + c_{M}\tilde{m}^3 + c_{MR}\tilde{m}^2\tilde{r} + c_{RM}\tilde{m}\tilde{r}^2 + c_{R}\tilde{r}^3 \label{eq:FundamentalPeriodFit} \\
    & + a_Z\log(Z) + a_Y Y + a_{\co} \log\biggl[\frac{\co}{(\co)_{\rm ref}}\biggr] \,, \nonumber
\end{align}
where $\tilde{m}=\log(M/\Msun)$ and $\tilde{r}=\log(R/\Rsun)$.
Best-fit coefficients are listed in Table~\ref{tab:CoefficientsFundamental}.
Predictions obtained with Eq.~\ref{eq:FundamentalPeriodFit}
generally match the models periods to within $\lesssim20$ per cent,
except for the high-mass and long-period regime,
where significant deviations occur. The residuals
from the fit are displayed in Fig.~\ref{fig:FitFundamentalPeriod}.
Note that the fit is limited to models for which
$\partial\log(\Teff)/\partial\log(L)<0$, \ie,
models that are cooling down as luminosity increases.
The models with the largest positive deviations
(around $\log(P)\sim3$) are those for which the structure is
already significantly different from a fully-convective
configuration (they are beginning to depart from
the Hayashi line). These models are probably poorly representative
of real pulsating stars. The histogram on top of
the residuals plot in Fig.~\ref{fig:FitFundamentalPeriod} shows how
the largest majority of models have their periods
reproduced to within $\sim10$ per cent.

\begin{table}
\normalsize
\centering
\caption{
    Coefficients of Eq.~\ref{eq:logPMR} for overtone modes periods.
    }
\label{tab:PMRcoefficientsOvertones}
\begin{tabular}{c||c|c|c|c}
 $n$ & 1 & 2 & 3 & 4 \\
\hline
$a_0$       & -1.55430 & -1.59749 & -1.74684 & -1.84377 \\
$a_M$       & -0.52869 & -0.48246 & -0.52692 & -0.52069 \\
$a_R$       & +1.57025 & +1.49958 & +1.52273 & +1.52656 \\
\end{tabular}
%\end{table}

%\begin{table}
%\normalsize
%\centering
\caption{
    Coefficients of Eq.~\ref{eq:FundamentalPeriodFit} for the fundamental mode period.
    }
\label{tab:CoefficientsFundamental}
\begin{tabular}{c|c|c|c}
\hline
$a_0$    & -1.12166 & $c_{M}$   & -0.07659 \\
$a_M$    & +1.24449 & $c_{MR}$  & -0.26130 \\
$a_R$    & +1.07886 & $c_{RM}$  & +0.26867 \\
$b_{M}$  & +0.87741 & $c_{R}$   & +0.03278 \\
$b_{MR}$ & -1.53239 & $a_Z$     & -0.02713 \\
$b_{R}$  & +0.10382 & $a_Y$     & +0.14872 \\
         &          & $a_{\co}$ & -0.01455 \\
\end{tabular}
\end{table}

\begin{figure*}
    \includegraphics[width=.75\textwidth]{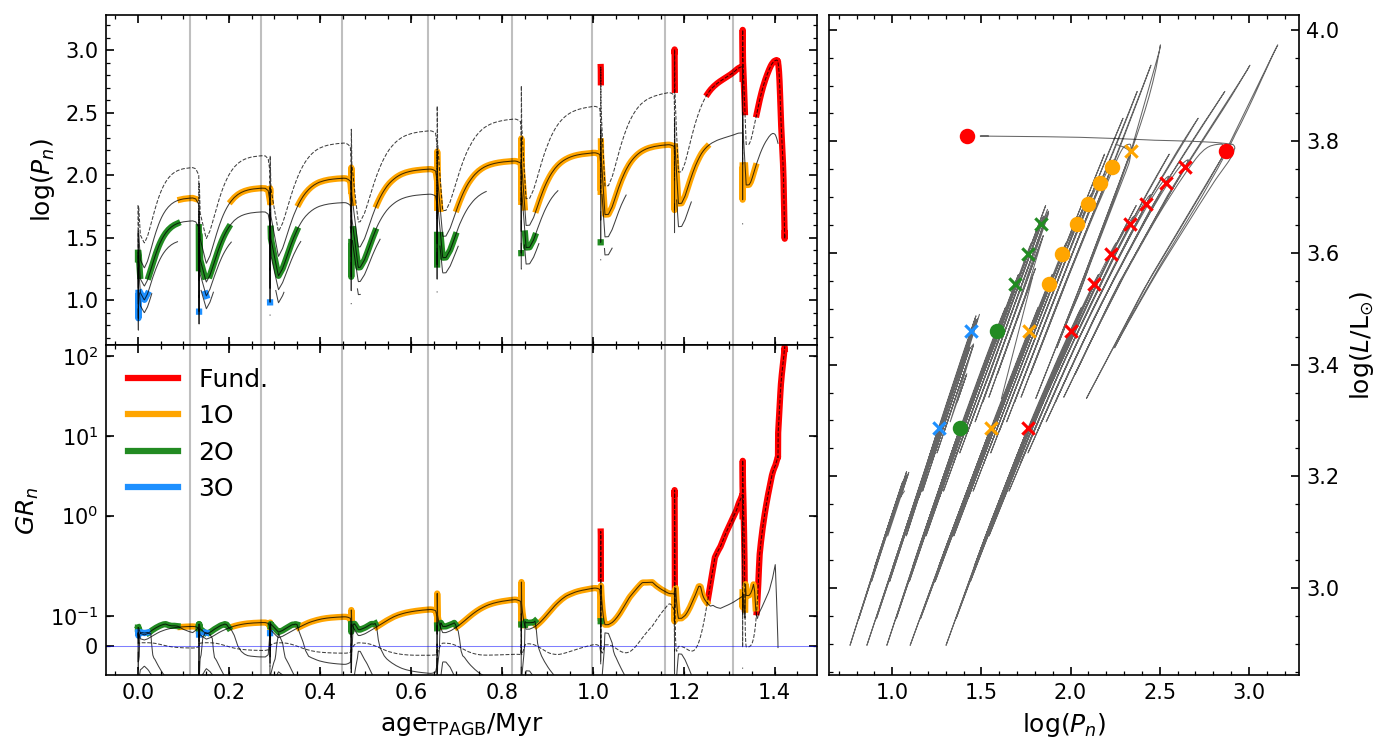}
    \caption{Left panels: periods (top) and growth rates (bottom)
    as a function of time elapsed since the beginning of TP-AGB
    in a \texttt{COLIBRI} evolutionary track with $M_0=1.5\,\Msun$
    and $Z_0=0.008$ at the beginning of the AGB. Dominant modes
    are highlighted by thick solid lines in colours.
    Modes other than the dominant are shown as thin solid lines,
    except for the fundamental that is shown as dashed thin lines
    to be more easily distinguishable.
    Vertical lines mark the point of maximum luminosity of
    quiescent evolution at each thermal pulse cycle.
    Note that the 4O mode is never dominant.
    Right panel: theoretical period-luminosity diagram.
    Symbols correspond to quiescent evolutionary points,
    with the dominant mode represented by a filled circle.
    }
    \label{fig:PLD_age_EvoTrack_M15}
\end{figure*}

\begin{figure*}
    \includegraphics[width=.75\textwidth]{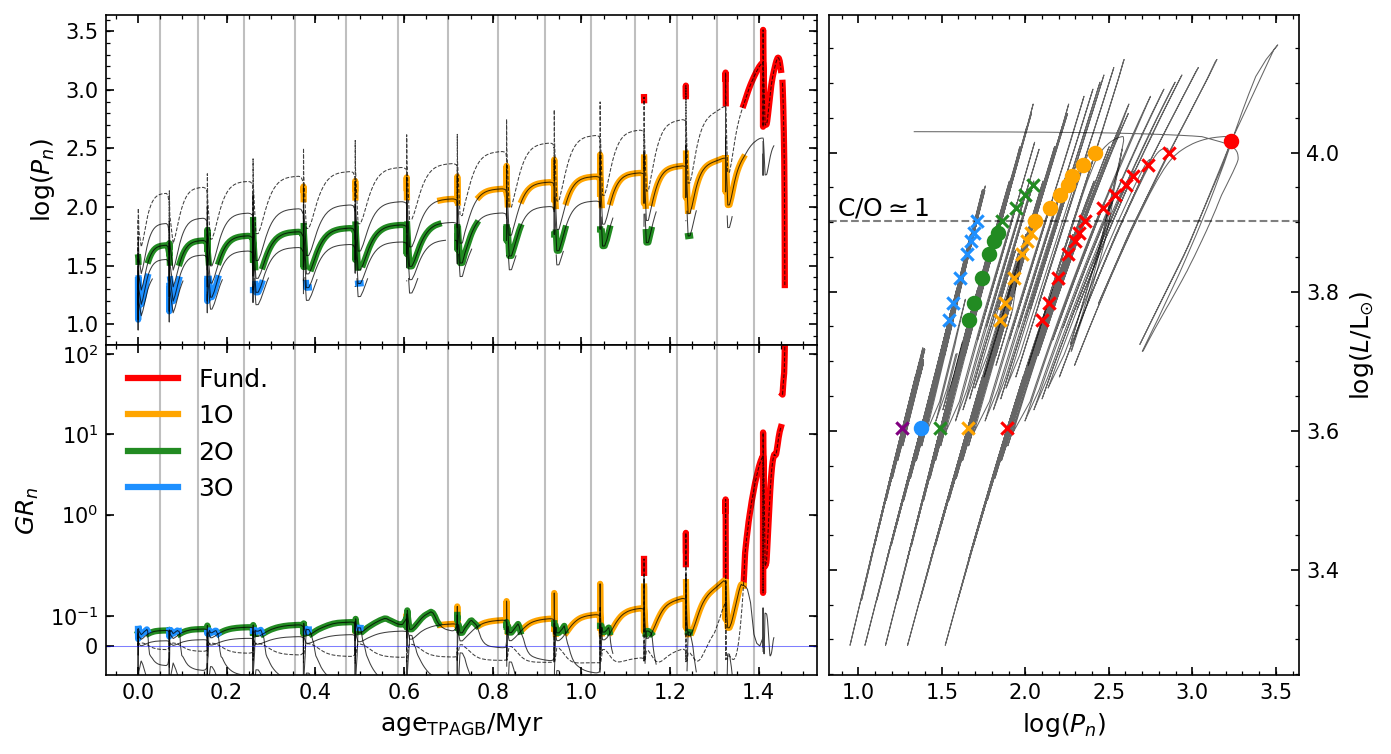}
    \caption{Same as Fig.~\ref{fig:PLD_age_EvoTrack_M15} but for
    a \texttt{COLIBRI} evolutionary track with $M_0=2.6\,\Msun$
    and $Z_0=0.008$ at the beginning of the AGB.}
    \label{fig:PLD_age_EvoTrack_M26}
\end{figure*}

\begin{figure*}
    \includegraphics[width=.75\textwidth]{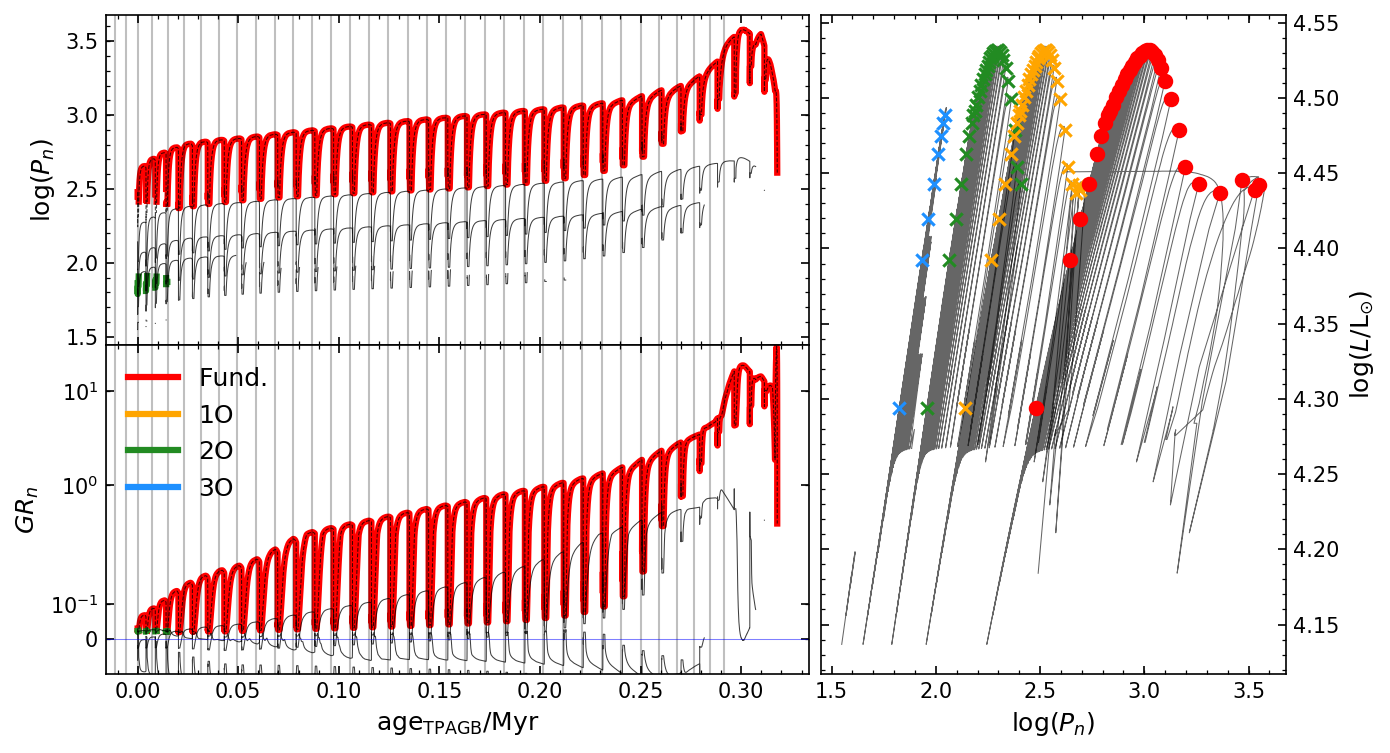}
    \caption{Same as Fig.~\ref{fig:PLD_age_EvoTrack_M15} but for
    a \texttt{COLIBRI} evolutionary track with $M_0=4.8\,\Msun$
    and $Z_0=0.008$ at the beginning of the AGB.}
    \label{fig:PLD_age_EvoTrack_M48}
\end{figure*}

\begin{figure*}
    \includegraphics[width=\textwidth]{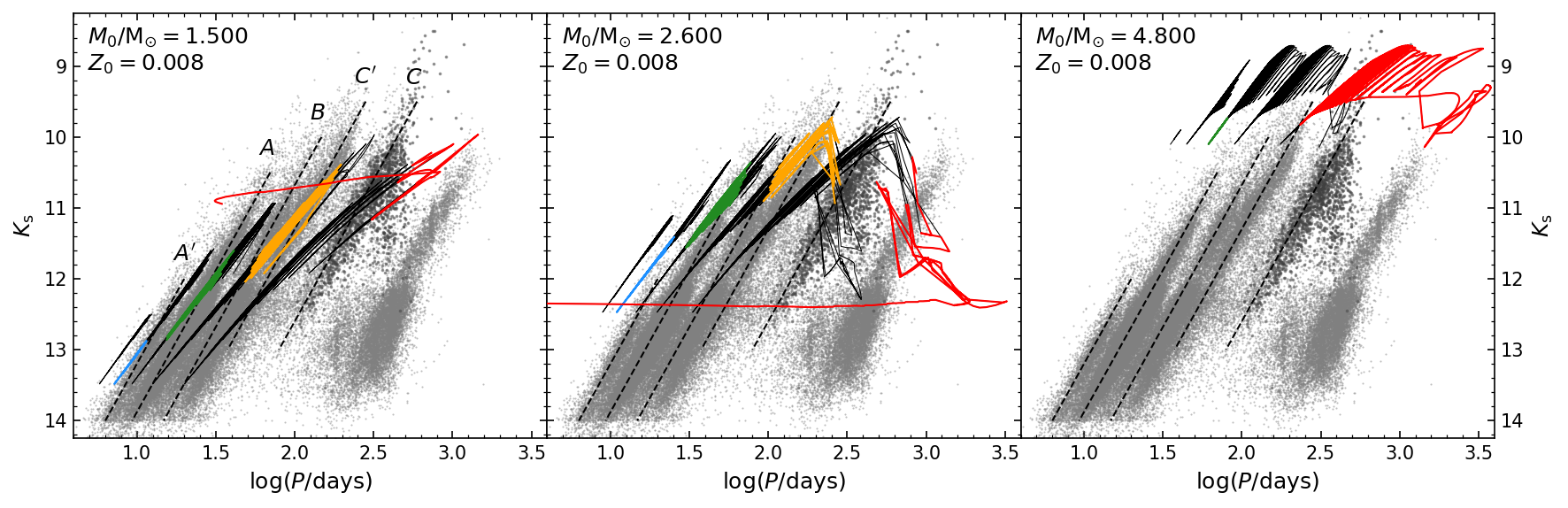}
    \caption{Evolutionary tracks from Figures~\ref{fig:PLD_age_EvoTrack_M15}, \ref{fig:PLD_age_EvoTrack_M26},
    and~\ref{fig:PLD_age_EvoTrack_M48} superimposed to the observed $\ks$--$\log(P)$
    diagram of observed LPVs in the LMC, from the OGLE-3 Catalogue \citep{Soszynski_etal_2009_LMC}.
    Dashed lines identify the position of observed PL sequences.
    Observed stars classified as Miras are shown with darker colours on sequence C.
    For each evolutionary track, the periods of five modes are shown with
    coloured solid lines where dominant, and black lines otherwise.}
    \label{fig:EvoTracks_OGLE3}
\end{figure*}

\section{Pulsation along evolutionary tracks}
\label{sec:PulsationEvoTrack}

In light of the simplified picture of pulsational
evolution described above, we examine the behaviour
of periods and growth rates for proper evolutionary models.
We consider three TP-AGB evolutionary tracks computed
with the \texttt{COLIBRI} code \citep{Marigo_etal_2013, Marigo_etal_2017},
with masses $M_0=1.5,\,2.6,$ and $4.8\,\Msun$, and $Z_0=0.008$,
where $M_0$ and $Z_0$ are the values of total mass and metallicity
at the beginning of the AGB phase. Periods and growth
rates, computed by interpolation in the grid of
pulsation models (see Appendix~\ref{sec:InterpolationRoutine})
are displayed in Fig.~\ref{fig:PLD_age_EvoTrack_M15}, \ref{fig:PLD_age_EvoTrack_M26}
and~\ref{fig:PLD_age_EvoTrack_M48} as a function of time elapsed since the beginning of TP-AGB.

By looking at the temporal evolution of periods
for the $1.5\,\Msun$ and $2.6\,\Msun$ tracks,
it is easy to recognise the stability pattern
described in Sect.~\ref{sec:StabilityOvertones},
with the dominant mode gradually shifting
towards lower radial orders\footnote{
    Although not explicitly depicted in the figures, other
    modes are excited beyond the dominant,
    in accordance with the fact that LPVs
    are multi-periodic, at least during part
    of their existence.}.
On the other hand, the presence of thermal
pulses makes hard to identify the familiar
pattern of growth rates of Fig.~\ref{fig:SelectedSequence},
thus justifying the simplified approach
employed in the previous sections.
In the right panels of Fig.~\ref{fig:PLD_age_EvoTrack_M15}, \ref{fig:PLD_age_EvoTrack_M26},
and~\ref{fig:PLD_age_EvoTrack_M48} we show the
theoretical period-luminosity diagram, highlighting
the location of quiescent evolutionary points.

Using evolutionary tracks, we can examine the
role of physical processes ignored in the
discussion in the previous sections,
such as mass-loss and the occurrence of thermal pulses.
The effect of mass-loss is particularly clear
in the PLD of Fig.~\ref{fig:PLD_age_EvoTrack_M15},
where we can see that the quiescent evolutionary
points bend towards longer periods as luminosity increases,
especially for the fundamental mode.
In contrast, the evolution of periods
with luminosity at constant mass would
follow an essentially straight line
in the $\log(P)$ vs. $\log(L)$ diagram,
similar to that depicted in Fig.~\ref{fig:SelectedSequence}.
In the PLD of Fig.~\ref{fig:PLD_age_EvoTrack_M26}
the bending is emphasised by the models
transitioning to C-rich, which causes an
increase of radius with respect to O-rich models
at the same luminosity. A change in slope
is particularly evident where $\co\simeq1$.
The $4.8\,\Msun$ evolutionary track in Fig.~\ref{fig:PLD_age_EvoTrack_M48}
represents a model undergoing Hot-Bottom Burning (HBB).
When this process is active, the stellar luminosity
is much higher than what predicted by a
CMLR in absence of HBB. By providing two
values of core mass in the grid of pulsation models,
we are able to account for this effect
in the estimation of periods and growth rates.

Finally, it is interesting to examine the
behaviour of pulsational stability with
respect to thermal pulses.
In the case of the $1.5\,\Msun$
track, the 1O mode is dominant during most of the quiescent
points (marked by vertical lines), with the exception of the last two thermal pulses
in which the fundamental mode becomes dominant.
In contrast, a substantial portion the $2.6\,\Msun$
track is dominated by 2O mode pulsation.
As discussed in Sect.~\ref{sec:DependenceGrowthRatesStellarParameters},
the fundamental mode is virtually always excited
and dominant for the $4.8\,\Msun$ evolutionary track.
Interestingly, the dominant
mode is not the same during individual thermal pulse cycles (TPCs).
This behaviour is largely
determined by the fact that the star shrinks immediately
following a thermal pulse. The decrease in radius causes
all pulsation frequencies, as well as the acoustic cut-off frequency (see Eq.~\ref{eq:nuac_isothermal}),
to increase, so that modes that had become stable by exceeding
the cut-off are brought back to an unstable configuration.
This can be easily understood by imagining that,
during the immediate post-flash, models travel backwards
along the sequence depicted in Fig.~\ref{fig:SelectedSequence}.

It is instructive to compare the evolutionary tracks
discussed above with observations, a shown in Fig.~\ref{fig:EvoTracks_OGLE3}.
We use variability data for LPVs in the LMC from the
OGLE-III database \cite{Soszynski_etal_2009_LMC}
combined with $JHK_{\rm s}$ photometry from the Two Micron
All Sky Survey \citep[2MASS;][]{Skrutskie_etal_2006}.
The comparison is made in the $\ks$--$\log(P)$ diagram,
where observed period-luminosity sequences are most clear.
In the left panel of Fig.~\ref{fig:EvoTracks_OGLE3} we show
the evolution of the $1.5\,\Msun$ track. When its
luminosity is low, at the beginning of the TP-AGB, the 3O
mode is dominant, with periods on the left edge of sequence A$^{\prime}$.
Upon reaching the acoustic cut-off frequency, the 3O mode becomes stable,
and the 2O mode becomes dominant on sequence A.
Later, the 1O mode becomes dominant, with periods
crossing sequences B and C$^{\prime}$, and finally
the fundamental mode becomes dominant on sequence C.
Note, however, that individual modes are not limited
to a single sequence \citep[see][and references therein]{Trabucchi_etal_2017}.
In fact, the 2O mode is excited
(although not dominant) on sequence A$^{\prime}$ too.
Similarly, the 1O mode is excited also on sequence A,
and the fundamental mode is found on sequences B and C$^{\prime}$ as well.

The $2.6\,\Msun$ evolutionary track shows
a qualitatively similar evolution, with the same
correspondence between dominant modes and sequences,
but at higher luminosities and longer periods
(the variation of mass throughout the PL sequences
in the LMC was discussed from an observational point
of view in \citet{Wood_2015}).
In this track, as well as the $1.5\,\Msun$ one,
the fundamental mode is predicted to
be dominant with periods longer than those
observed on sequence C. This is a direct
consequence of the regime of temporary
stabilisation described in Sect.~\ref{sec:StabilityFundamental}.
This disagreement with observations is also
discussed in \citet{Trabucchi_etal_2017}.
A detailed analysis of this aspect is
beyond the scope of the present work,
and will be addressed in a forthcoming paper.

The $4.8\,\Msun$ track appears to populate only
sequence C, in accordance with the prediction
that the fundamental mode is always dominant.
Its location is in agreement with that of
the brightest Miras in the observed sample.
Note that the scarcity of observed stars in that
region is compatible with the brief lifetimes
of these objects, indeed the $4.8\,\Msun$ evolutionary
track has a duration five times shorter than that of the
other two (cf. Fig.~\ref{fig:PLD_age_EvoTrack_M15},
\ref{fig:PLD_age_EvoTrack_M26}, and~\ref{fig:PLD_age_EvoTrack_M48}).

\section{Conclusions}
\label{sec:Conclusions}

We presented a new grid of linear, non-adiabatic, radial pulsation models
for red giant stars, widely covering the space of stellar parameters
characteristic of the TP-AGB evolutionary phase. Models include
periods and amplitude growth rates for five oscillation modes,
from the fourth overtone to the fundamental mode. The grid is
made public through a dedicated web interface, the
details of which are presented in Appendix~\ref{sec:WebInterface}, and represents a significant
update with respect to previously published models of long-period variables.
The main novelty is the exploration of a wide space of chemical
compositions in terms of metallicity and $\co$ ratio, coupled
with detailed atomic and molecular opacities consistent with
the specific metal mixture assumed for the envelope.
This allows for a consistent description of stellar
pulsation in stars that experience significant alterations of
the envelope composition due to repeated third dredge-up events,
and makes the present set of models the first one to
systematically account for variability in C-type stars.

We employed models from our grid to provide a general
picture of the evolution of pulsation in long-period variables,
the main results being the following.
\begin{enumerate}

\item   The period of overtone modes is determined by global
        properties mass and radius, and is weakly dependent
        on other parameters or on the interior structure.
        Their growth rates are more sensitive than periods to other quantities,
        but have the interesting properties of dropping to
        negative values when oscillation frequencies
        exceed the acoustic cut-off, largely determined
        by surface properties.

\item   As a result, overtone modes follow a rather
        regular pattern, in which the dominant pulsation mode
        shifts towards lower radial orders as the models expand, 
        and the number of expected excited modes decreases.
        This tendency is similar to that presented by solar-like oscillation spectra of less evolved red giants.

\item   In contrast, the properties of the fundamental mode
        are strongly affected by the internal structure of the models,
        and in particular by the displacement of the partial
        ionization zones of major elements associated to the envelope expansion.
        For this reason, it is much harder to provide a general
        picture of the fundamental mode as a function of global
        stellar parameters. For relatively low mass models ($\lesssim3\,\Msun$),
        the fundamental mode exhibits a regime of temporary
        stabilisation followed by rapid rise in growth rates,
        and becomes dominant at large radii just after
        the 1O mode exceeds the acoustic cut-off.
        In more massive models, on the other hand, it
        appears to be almost always excited and dominant.

\item   The period of the fundamental mode is mainly determined by mass and radius,
        a dependence that can be expressed
        to a first approximation as $P\propto M^{a_M}R^{a_R}$.
        However, the power-law exponents $a_M$ and $a_R$ depend
        themselves upon mass and radius, in a somewhat erratic fashion.
        As a consequence, models with different composition
        and input physics show possibly large deviations
        from a simple period-mass-radius relation.
        These properties make it very difficult to derive
        a best-fit relation for the fundamental
        mode period that is both simple and accurate.

\item   We examined in more detail the evolution of pulsation
        during the TP-AGB by computing periods and growth
        rates for a few evolutionary tracks, and
        found the results to be consistent with observations
        of long-period variables in the Large Magellanic Cloud.
        The shift of the dominant mode of pulsation 
        towards lower radial orders as stars climb the AGB
        can be easily put in relation to the
        observed period-luminosity sequences.
        The radial order of the dominant mode
        can also change during individual thermal pulse cycles,
        as a consequence of the temporary shrinking of the envelope.
        Pulsation periods, especially for the fundamental mode,
        are significantly affected by mass loss and by the
        transition from O-rich to C-rich composition.
        In both cases, the effect is mainly driven by
        differences in mass and radius at given luminosity.

\item   By exploiting the regularity of overtone modes, we provided best-fit
        relations for their periods, and a simple
        prescription to predict the most-likely dominant
        overtone in a stellar model as a function of global
        stellar parameters. For not-too-massive models
        ($M\lesssim 3\Msun$), the
        fundamental mode fits this scheme reasonably well.
        A best-fit relation for the fundamental mode period
        is also provided, including the dependence upon chemical composition.
\end{enumerate}

We stress that analytic relations such as those provided here
cannot possibly grasp the full complexity of pulsation and its
dependence on the properties of stellar models.
Therefore, they cannot be used to make accurate
predictions, a task for which we encourage the use
of the full grid of models and the companion interpolation
routine provided via the web interface.

As a next step to expand the present work, we plan
to include additional values of metallicity to the
grid, and to explore the effect of non scaled-solar
oxygen abundances.

In subsequent works, the models computed here
will be used to study the pulsation properties,
especially the PL sequences, of the populations
of red giants in stellar systems.
Of particular importance are the LMC, SMC,
and the Galactic Bulge where large numbers
of red giant variables are known from
the OGLE and MACHO surveys, as well as
other systems such as globular clusters
and local dwarf galaxies.

\section*{Acknowledgements}

We acknowledge the support from the ERC Consolidator Grant funding scheme
({\em project STARKEY}, G.A. n. 615604).
This publication makes use of data products from the Two Micron All Sky Survey,
which is a joint project of the University of Massachusetts
and the Infrared Processing and Analysis Center/California Institute of Technology,
funded by the National Aeronautics and Space Administration and the National Science Foundation.
This publication makes use of data from the OGLE-III Catalog of Variable Stars.

%%%%%%%%%%%%%%%%%%%%%%%%%%%%%%%%%%%%%%%%%%%%%%%%%%

%%%%%%%%%%%%%%%%%%%% REFERENCES %%%%%%%%%%%%%%%%%%

\bibliographystyle{mnras}
\bibliography{References}

%%%%%%%%%%%%%%%%%%%%%%%%%%%%%%%%%%%%%%%%%%%%%%%%%%

%%%%%%%%%%%%%%%%% APPENDICES %%%%%%%%%%%%%%%%%%%%%

\appendix

\section{Web Interface}
\label{sec:WebInterface}

The pulsation models can be accessed via the web interface
hosted on the web site \url{http://starkey.astro.unipd.it/web/guest/pulsation}.
The interface allows the computation of pulsation properties
for arbitrary combinations of global stellar parameters
by linear interpolation in the grid of models (see \ref{sec:InterpolationRoutine}).
The web interface includes the possibility to manually enter the
stellar parameters (for individual stellar moderls)
or to upload a properly formatted file containing
a tabulated list of the required combinations (for isochrones,
evolutionary tracks, or synthetic stellar population models).
The whole grid of models, together with
the companion interpolation program, is also available for download from the website.

\subsection{Interpolation Routine}
\label{sec:InterpolationRoutine}

The interpolation routine allows the computation of the
pulsation properties for a given stellar model
represented by a combination of
seven global stellar parameters ($Z$, $X$, $\co$,
$M/\Msun$, $\Mc/\Msun$, $\log(\Teff)$, and $\log(L/\Lsun)$),
that are defined by the user input.
Each one of these parameters correspond to
one of the dimensions of the grid.
The grid is regular in $\Zref$, $X$, $\co$, and $M$,
so that the appropriate interpolation nodes can be
immediately identified based on the input values.
The models are indexed according to their
reference metallicity $\Zref$, but interpolation
is performed over the value of `true' metallicity
(that is stored in the models files).
Thus, the input value has to be that of the
`current' metallicity. The corresponding value of $\Zref$
(used to identify the appropriate interpolation nodes)
is computed from $Z$ and $\co$ as
\begin{equation}\label{eq:Apdx_Zref}
    \Zref = Z \left\{1 + \left(\frac{\xc}{Z}\right)_{\rm ref} \left[\frac{\co}{(\co)_{\rm ref}} - 1\right] \right\}^{-1} \,,
\end{equation}
\ie, under the assumption that changes in $Z$
and $\co$ occur only as a consequence of variations
in the abundance of carbon (consistently with the
assumptions made in the construction of the grid of models,
see Sect.~\ref{sec:CoverageStellarParameters}).
The grid is not exactly regular in core mass,
which is related to luminosity by Eq.~\ref{eq:CoreMassLuminosityFunctions},
so that nodes do not have fixed values.
However this is not an issue since exactly two
$M_c$ nodes are available, and thus both used.
The most problematic parameter is the $\Teff$,
which value is determined by envelope integration
for a given set of input parameters and a given
value if $\aml$, thus cannot be known a priori from
other parameters. To account for this, the interpolation
proceeds in a series of successive steps, as follows.

First, the value of $\Zref$ corresponding to the input $Z$
is computed, and the appropriate interpolation nodes are
identified, as well as thus for $X$, $M$, and $\co$.
The resulting four pairs of nodes, plus two nodes of $\Mc$
and three nodes of $\aml$, correspond to $3\times2^5=96$
combinations, each one identifying a sequence of models
with increasing luminosity. For each series of models,
the nodes of luminosity bracketing the input value $\log(L_{\rm IN})$
are identified, and linear interpolation over those
nodes is performed for all periods and growth rates,
as well as for $\Mc$ and $\log(\Teff)$. The values of $\Mc$
computed this way are then used as interpolation nodes
in that dimension of the grid. The same operation is
then performed for $M$, $\co$, $X$, and $Z$ in this order,
At each step, a new value of $\log(\Teff)$ is also computed.
In the end, one is left with three values of $\log(\Teff)$,
corresponding to the three $\aml$ nodes, and the associated
values of periods and growth rates, resulting from
the successive interpolation. The two nodes of effective
temperature bracketing the input value are identified,
and a final linear interpolation is performed.

Note that the interpolation routine allows for
extrapolation wherever the input value is outside
the grid boundaries, which are often not known a priori
(\eg, for the $\Teff$, or for $\log(L)$, where the boundaries
depends upon mass and chemical composition).
Occurrences of extrapolation are recorded in
a flag variable that is returned together with
the interpolated values of periods and growth rates.
The flag distinguishes between extrapolation
`upwards' or `downwards'. Estimates of pulsation
properties for which any extrapolation occurred
should be treated with some care. For typical
input values, it is not infrequent to obtain
extrapolation to lower masses, higher metallicity,
or higher or lower values of core mass, luminosity and $\Teff$.
A number of tests suggest that extrapolated values are
generally safe (provided the input values are not
too extreme), with the exception of extrapolation
towards luminosity higher than the grid limits,
in which case the result is likely unreliable.

\begin{figure}
    \includegraphics[width=\columnwidth]{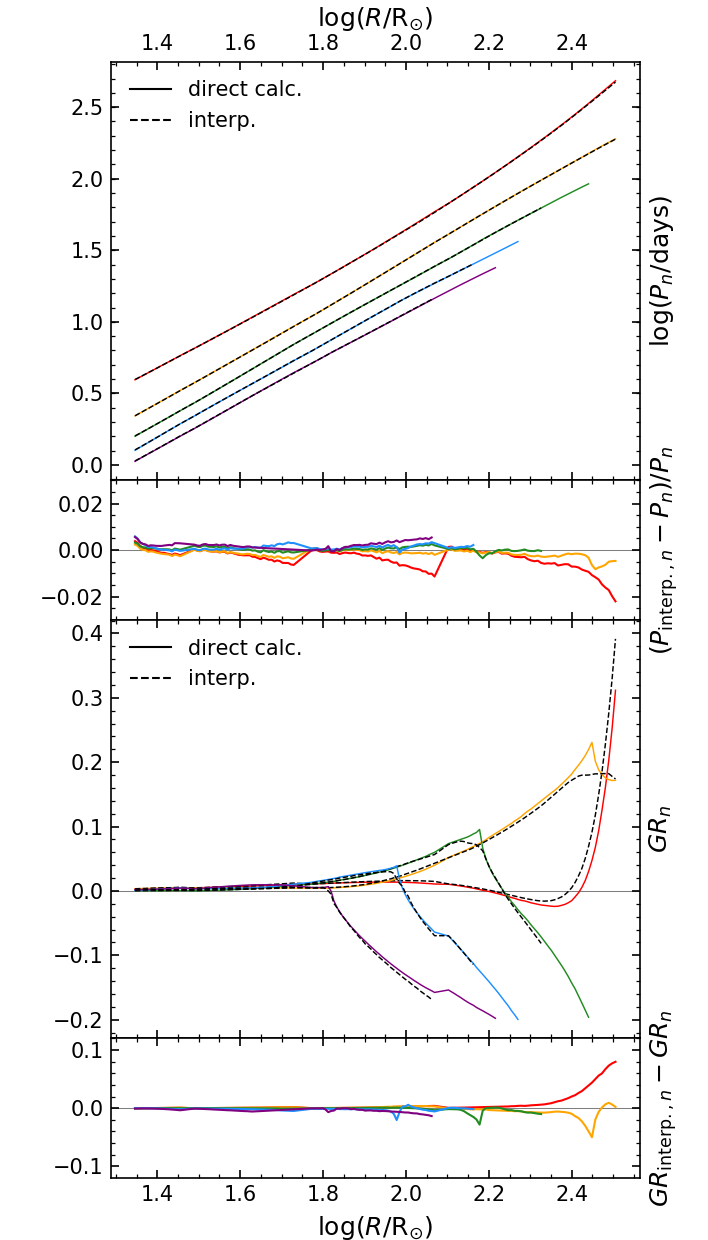}
    \caption{Comparison between periods and growth rates obtained by
    direct calculation and by linear interpolation over the grid of
    pulsation models. Periods are essentially indistinguishable between
    the two approaches. Although differences are visible for growth rates,
    results from the interpolation reproduce qualitatively well those
    from direct computation.}
    \label{fig:interp_valid}
\end{figure}

A simple way to test the interpolation scheme
is to attempt to recover the results of direct computation.
An example of this test is shown in Fig.~\ref{fig:interp_valid}.
There, we display the periods and growth rates
of a series of models specifically computed in
which all parameters corresponding to grid nodes
vary continuously and monotonically, keeping
values within the boundaries of the grid.
The corresponding values of periods and growth
rates, computed by interpolation over the grid
for each model of the series, are also shown there,
as well as the differences between the two cases.

Periods are mostly recovered to within less than $1$ per cent,
a result largely due to the fact that $\log(P_n)$
shows a monotonic dependence upon all parameters,
and that dependence is rather close to linear (locally).
Growth rates, on the other hand, do not depend
monotonically upon global stellar parameters, leading
to less accurate results from the interpolation.
This is especially true in the neighbourhood of
the acoustic cut-off, where growth rates form a cusp
that is displaced in luminosity (radius) for
different values of mass and chemical composition
(as discussed in Sect.~\ref{sec:DependenceGrowthRatesStellarParameters}).

%%%%%%%%%%%%%%%%%%%%%%%%%%%%%%%%%%%%%%%%%%%%%%%%%%

% Don't change these lines
\bsp	% typesetting comment
\label{lastpage}
\end{document}